\documentclass[12pt,a4paper,final]{iopart}
\usepackage{iopams}  
\usepackage{graphicx}
\usepackage[breaklinks=true,colorlinks=true,linkcolor=blue,urlcolor=blue,citecolor=blue]{hyperref}

\expandafter\let\csname equation*\endcsname\relax 
\expandafter\let\csname endequation*\endcsname\relax

\usepackage[utf8]{inputenc}
\usepackage[english]{babel}
\usepackage{array, booktabs}
\usepackage[colorinlistoftodos]{todonotes}
\usepackage{nameref}
\usepackage{amsthm,amsmath,amssymb,mathtools,epsfig,makecell,listings, paralist, enumitem} %
\usepackage[font=small, labelfont=bf]{caption}
\usepackage[font=footnotesize, labelfont=bf]{subcaption}
\usepackage{changepage}
\usepackage{fancybox}
\usepackage{todonotes}
\usepackage{menukeys}
\usepackage{rotating}

\newcommand{\R}{\ensuremath{\mathbb{R}}}
\newcounter{todocounter}

\presetkeys{todonotes}{inline, color=blue!30}{}

\usepackage{cite}

\begin{document}

\title[TDA of Task-Based Schizophrenia fMRI Data]{Topological Data Analysis of Task-Based fMRI Data from Experiments on Schizophrenia}

\author{Bernadette J. Stolz}
\address{Mathematical Institute, University of Oxford, Oxford, UK}
\ead{stolz@maths.ox.ac.uk}

\author{Tegan Emerson}
\address{Pacific Northwest National Laboratory, Seattle, WA USA}
\ead{tegan.emerson@pnnl.gov}

\author[cor1]{Satu Nahkuri}
\address{F. Hoffmann-La Roche AG, Basel, Switzerland}
\ead{satu.nahkuri@roche.com}

\author{Mason A. Porter$^{1,2}$}
\address{$^1$Department of Mathematics, University of California, Los Angeles, Los Angeles, USA}
\address{$^2$Mathematical Institute, University of Oxford, Oxford, UK}
\ead{mason@math.ucla.edu}

\author{Heather A. Harrington}
\address{Mathematical Institute, University of Oxford, Oxford, UK}
\ead{harrington@maths.ox.ac.uk}

\begin{abstract}
We use methods from computational algebraic topology to study functional brain networks, in which nodes represent brain regions and weighted edges encode the similarity of fMRI time series from each region. With these tools, which allow one to characterize topological invariants such as loops in high-dimensional data, we are able to gain understanding into low-dimensional structures in networks in a way that complements traditional approaches that are based on pairwise interactions. In the present paper, we use persistent homology to analyze networks that we construct from task-based fMRI data from schizophrenia patients, healthy controls, and healthy siblings of schizophrenia patients. We thereby explore the persistence of topological structures such as loops at different scales in these networks. We use persistence landscapes and persistence images to create output summaries from our persistent-homology calculations, and we study the persistence landscapes and images using $k$-means clustering and community detection. Based on our analysis of persistence landscapes, we find that the members of the sibling cohort have topological features (specifically, their 1-dimensional loops) that are distinct from the other two cohorts. From the persistence images, we are able to distinguish all three subject groups and to determine the brain regions in the loops (with four or more edges) that allow us to make these distinctions.
\end{abstract}

\vspace{2pc}




\section{Introduction}

Schizophrenia is a chronic psychiatric disorder that affects more than 21 million people worldwide~\cite{WHOSchizophrenia}. Up to 80\% of the risk factors appear to be genetic, although it has proven difficult to identify the specific genes that are involved in the disease~\cite{Bertolino2009II}. The disease usually commences in early adulthood, and symptoms range from hallucinations and avolition to cognitive deficits (such as impaired working memory)~\cite{WHOSchizophrenia,Dawson2014}. It is believed that the cognitive deficits arise from compromised functional integration between neural subsystems~\cite{Bullmore1997,Peled1999, Bassett2008,Dawson2014}. There can be significant differences in the properties of time series from imaging measurements of healthy versus schizophrenic individuals, although different studies have found seemingly contradictory results when comparing functional magnetic resonance imaging (fMRI) time series from two distinct brain regions in a schizophrenia patient and a healthy control. The majority of studies have concluded that schizophrenia patients have less-similar time series than healthy controls across different brain regions 
\cite{Fornito2012}. Zalesky \emph{et al.}~\cite{Zalesky2012} suggested that such reduced similarity may arise from an altered coupling between brain regions and local decoherence within brain regions in schizophrenia patients. However, some studies have observed that schizophrenia patients have more-similar series than healthy controls across brain regions. For a detailed discussion of these seemingly contradictory findings, see~\cite{Fornito2015}. In some cases, methodological steps in fMRI analyses seem to yield increases in these similarities, but abnormal neurodevelopment or drug treatment may play a role in increasing them in other cases~\cite{Fornito2015}.

\begin{figure}[h!]
\centering\includegraphics[width=\textwidth]{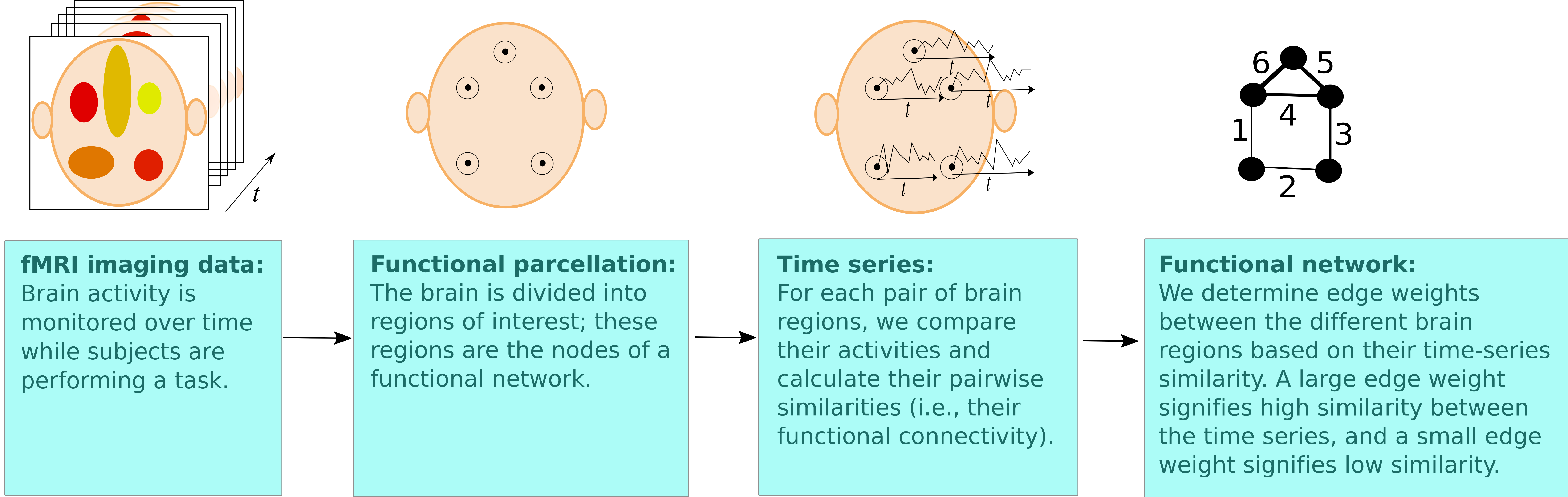}
\caption{Pipeline to construct functional networks from imaging data (e.g., fMRI data).
}
\label{fig:FunctionalNetwork}
\end{figure}

One approach for studying the human brain is to construct a (possibly time-dependent) neuronal network based on experimental data and then analyze the network's structure and dynamics to gain insights into its properties~\cite{Bullmore2009,Bullmore2011,Sporns2014,Papo2014,Papo2014II,Betzel2016,bassett2017,bassett2018}. 
One can form a so-called \emph{functional network}~\cite{Bullmore2011,Bullmore2009,Sporns2015,Petersen2015,Stolz2017}, in which each node represents a brain region and one weights the edges between them based on some measure of the similarity between the nodes' fMRI time series. In Fig.~\ref{fig:FunctionalNetwork}, we show a pipeline of how to construct a functional network from fMRI time-series data. Additionally, when interpreting functional networks in fMRI studies, it is very important to consider the cautionary notes in \cite{eklund2016}.

Studies of functional networks of schizophrenia patients have revealed that such networks differ significantly from the functional networks of healthy controls~\cite{Lynall2010,Rubinov2013,Alexander-Bloch2012,Bassett2008,Liu2008,Fornito2012,Singh2016}. For example, schizophrenia patients can have rather different community structure
than controls~\cite{Alexander-Bloch2012,flanagan2018}. 
For example, Alexander-Bloch \emph{et al.}~\cite{Alexander-Bloch2012} observed that a small subset of brain regions lead to significant differences in the community assignments in schizophrenia patients, whereas the communities for healthy subjects appear to be consistent with each other. Moreover, the maximum modularity of functional networks appears to be smaller for schizophrenia patients than for healthy controls~\cite{Alexander-Bloch2012, Alexander-Bloch2010}. Two recent papers, Flanagan \emph{et al.}~\cite{flanagan2018} and Towlson \emph{et al.}~\cite{towlson2019}, compared the network structures of schizophrenia patients and healthy controls under the effects of different drugs and a placebo.

An increasingly popular approach for the analysis of functional networks is to use ideas from computational algebraic topology, as these approaches allow one to characterize topological invariants (such as connectedness and loops) in high-dimensional data 
~\cite{Edelsbrunner2002,Edelsbrunner2008, Ghrist2008,Carlsson2009,Edelsbrunner2010,sizemore2018,batt2020}. In contrast to standard methods of network analysis \cite{Newman2018}, employing computational topology allows one to explicitly go beyond pairwise connections; this is helpful for gaining global understanding of low-dimensional structures in networks. Although one can also use frameworks such as hypergraphs~\cite{Bollobas1998} to study higher-order network structures (see, e.g., ~\cite{Bassett2014}), such a formalism does not by itself give direct information about the shape or scale of mesoscale features in networks. By contrast, \emph{persistent homology} (PH), the most prominent approach in topological data analysis, allows one to explore the persistence of features (such as connectedness and loops) in data sets \cite{otter2017,Patania2017}. Persistent homology has led to interesting insights in a variety of fields (for {examples, see~\cite{kramar2013,taylor2015,bgk2015,topaz2014,Bendich2014,Feng2020,Byrne2019})}; and it has been used increasingly in neuronal networks, leading to {several promising insights~\cite{Curto2008,Dabaghian2012,Petri2013,Lee2011,Giusti2015,Spreemann2015,Curto2016,Giusti2016,Reimann2017,Dabaghian2016,Stolz2017,Lee2019,Bardin2018,chung2019,babichev2018,geniesse2019,Ibanez2019}. }

In the present paper, we construct functional networks using fMRI data from schizophrenia patients, healthy controls, and siblings of schizophrenia patients. We create a nested sequence of networks in which we add edges, one by one, to the networks in order from the largest edge weights to the smallest. (In the unlikely case of two edges having the exact same weight, we add both edges simultaneously in one step.) We then construct a
 weight-rank clique filtration (WRCF) \cite{Petri2013} by determining cliques and tracking their changes in each step of the network sequence. 
 We then compute PH and Betti numbers~\cite{Edelsbrunner2010,Croom} of the WRCF and examine the results by applying tools from statistics and machine learning, respectively, on the persistence landscapes and persistence images that result from our computation of PH. We compare our findings from these two approaches. We focus on loops (with four or more edges) \footnote{We use the term `loop' to refer to at least four edges in a network that are connected in a way that forms a cycle. Conventionally, loops (other than self-loops) in undirected graphs must
have at least 3 edges, and loops in directed graphs must have at least 2
edges. In our paper, we adapt this terminology to represent the topological
features that we detect in our simplicial complexes.} in the networks in our nested sequence, rather than on connected components, because one can also study the latter using more conventional approaches (such as by computing the spectrum of the combinatorial graph Laplacian matrix \cite{Bollobas1998,Newman2018}). Although it is interesting to also consider higher-dimensional topological features in these functional networks, the computational cost of PH is very high~\cite{otter2017} (especially in higher dimensions), and we therefore focus on the analysis of loops.

Our paper proceeds as follows. We introduce the data set and the mathematical methods in an intuitive way in Section~\ref{Sec:MatMethods}, present our findings in Section~\ref{Sec:Results}, and discuss our comparisons in the context of current biological research in Section~\ref{Sec:Discussion}. We give some additional details about a few results and report our results of our computation of Betti curves in \ref{S1}.


\section{Methods}\label{Sec:MatMethods}

\subsection{Data set: {fMRI} data of schizophrenia patients, siblings of schizophrenia patients, and healthy controls} 

We use a data set that consists of time series from blood oxygen level-dependent (BOLD) functional magnetic resonance imaging (fMRI) data
that was collected from 281 subjects (encompassing 54 schizophrenia patients, 50 healthy siblings of schizophrenia patients, and 177 healthy controls) with 120 time steps (where the length of $1$ time step corresponds to $\Delta t = 2\, \text{s}$). The brain regions were determined according to the Montreal Neurological Institute template~\cite{Talairach1988}. Prior to obtaining the time series, the fMRI data were corrected for head motion, and they were normalized and smoothed with a Gaussian filter. The voxel-wise signal intensities were normalized to the whole-brain global mean. The data set was acquired by Bertolino, Blasi, and their collaborators as part of a larger fMRI data set over a period of approximately 10 years. Subsets of the data set have been studied previously \cite{Bertolino2010,Sambataro2009,Rampino2014}, although these previous studies of the data did not include the data for siblings. 

The experimentalists obtained fMRI images while subjects were performing a block paradigm of a so-called `$n$-back task'. 
During an $n$-back task, subjects are presented with a sequence of {stimuli (such as numbers)}. In each step $m$ of the sequence, subjects are first shown a number and then asked to recall the number from sequence step $m-n$. For example, during a $2$-back task, subjects are shown a sequence 
$\{\dots, x_{i-1}, x_i, x_{i+1}, x_{i+2}, \dots\}$ and are asked to recall number $x_{i-1}$ while being shown number $x_{i+1}$, recall number $x_i$ while being shown number $x_{i+2}$, and so on. For the present data set, the {block paradigm} consisted of alternating blocks of {four} $0$-back tasks and {four} $2$-back tasks. 

We preprocess the data to remove noise that arises due to (1) contributions from brain white matter~\cite{Weissenbacher2009} and cerebrospinal fluid~\cite{Weissenbacher2009,Dagli1999} (in these areas, one does not expect a response that is related to neuronal processes), (2) spontaneous global signal fluctuations~\cite{Weissenbacher2009,Birn2006,Fox2007}, and (3) signal mismatches between images from the head motion of subjects~\cite{Friston1996}. For each subject and time step, we calculate the mean signal for white-matter brain regions, the mean signal for regions that consist of cerebrospinal fluid, and the mean of the global signal. In addition to these mean values, we also include the squares and cubes of the global signal means, as well as head-motion parameters (3 translation and 3 rotation parameters), as rows in our $11 \times 120$ subject-specific design matrices. We then perform linear regression for each time series using {\sc Matlab}'s command for the Moore--Penrose pseudoinverse \footnote{Because some of the matrices are ill-conditioned, the resulting networks differ across different runs of the preprocessing. However, in our observations, the matrices differ by only up to $0.2\%$ of the entries between two separate realizations of preprocessing.}  {\sc pinv()}; we exclude brain regions without grey matter from our calculations. We then use the residuals from the regression as our time series for the $120$ brain regions that we list in Tables \ref{S1_Table: BrainRegionsI}--\ref{S1_Table: BrainRegionsV}.
Such preprocessing steps are common when working with fMRI data, but they are not uncontroversial. In particular, the effects of global signal regression can alter correlations between time series. See, for example, \cite{Murphy2009,Fox2009} and~\cite{Fornito2015} in the context of schizophrenia.


\subsection{Functional connectivity}\label{construct}

We construct functional networks from the fMRI time series for each subject by using the 120 distinct brain regions (see Tables \ref{S1_Table: BrainRegionsI}--\ref{S1_Table: BrainRegionsV}) as the nodes of the networks and calculating Pearson correlations \footnote{There are numerous ways to measure functional connectivity \cite{Smith2011,Zhou2009,Bullmore2011}. For a discussion in the context of schizophrenia research, see \cite{Fornito2012}.} (without a time lag) between the nodes' time series as a measure of pairwise functional connectivity. The values of the pairwise functional connectivity give the edge weights between the brain regions in the functional networks. 

\begin{figure}[h!]
\centering\includegraphics[width=\textwidth]{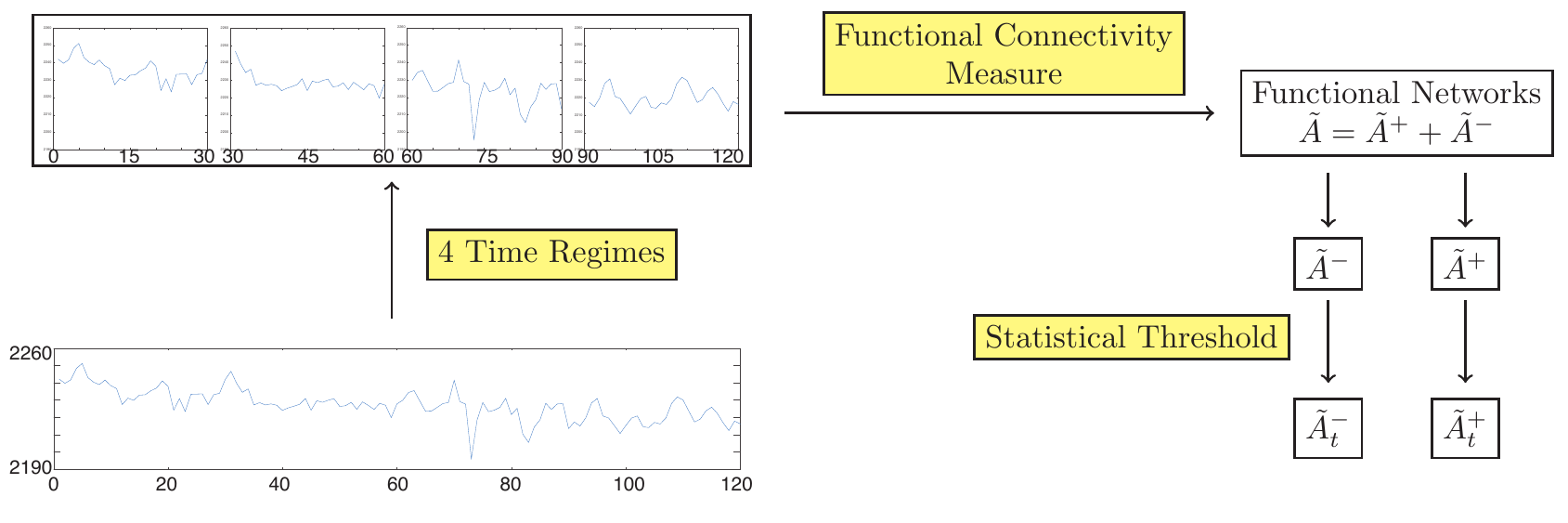}
\caption{Steps that we perform on the preprocessed time series of each brain region 
to construct a functional network for each subject during each of four time regimes. We study the positive parts of the resulting networks using persistent homology.
}\label{fig:TimeSeries}
\end{figure}

In our key computations in the present paper, we consider four contiguous time regimes of 30 time points each; this yields four functional networks per subject. (The lone exception to this approach is Section~\ref{S1_3}, in which we use each subject's full time series, which consists of 120 time points, to construct a single functional network for each subject.) Although the four time regimes {each overlap temporally with times during which subjects performed one $0$-back and one $2$-back task}, our separation into time regimes is motivated by an interest in potential developments in the dynamics over time, rather than in relating the fMRI response to the task. We represent each functional network using an adjacency matrix $A = A(\text{subject}, \text{time regime})$, whose entry $A_{ij}$ is given by the edge weight between node $i$ and node $j$. Due to the high computational cost of PH~\cite{otter2017}, we reduce the number of edges in the networks that we analyze to enable computations. We apply a statistical threshold, described in \cite{Bassett2011}, to the weighted adjacency matrices without modifying the remaining edge weights. To obtain the thresholded adjacency matrices, we estimate p-values for the correlations using the {\sc Matlab} function {\sc corrcoef} and retain only those entries whose p-value is less than $0.05$. Using this type of thresholding, we retain at most 44\% of the edges in a network and retain {a mean} of 20--30\% of the edges {in each subject group}. We then separate each adjacency matrix into a positive and a negative part, $A = A^+ + A^-$, and study only the positive $A^+ $ part of the adjacency matrix \footnote{We choose to only include positive edge weights to avoid the need to interpret negative correlations between time series.}. In discarding negative edge weights, we retain a mean of slightly more than 50\% of the entries in our thresholded adjacency matrices. In Fig.~\ref{fig:TimeSeries}, we show a diagram of the steps that we perform to construct our functional networks. Although we consider the four time regimes separately, we treat all subjects and all time regimes together as one data set.

\subsection{Persistent homology}

\emph{Persistent homology} (PH) is a technique from topological data analysis, which aims to understand the `shape' of data \cite{otter2017}. PH is based on the topological concept of \emph{homology}, which is used to study the shape of objects in a way that disregards changes from stretching and bending. 

\begin{figure}[h!]
\centering
\subcaptionbox{Swiss (Emmental) Cheese}{\includegraphics[width=0.49\textwidth]{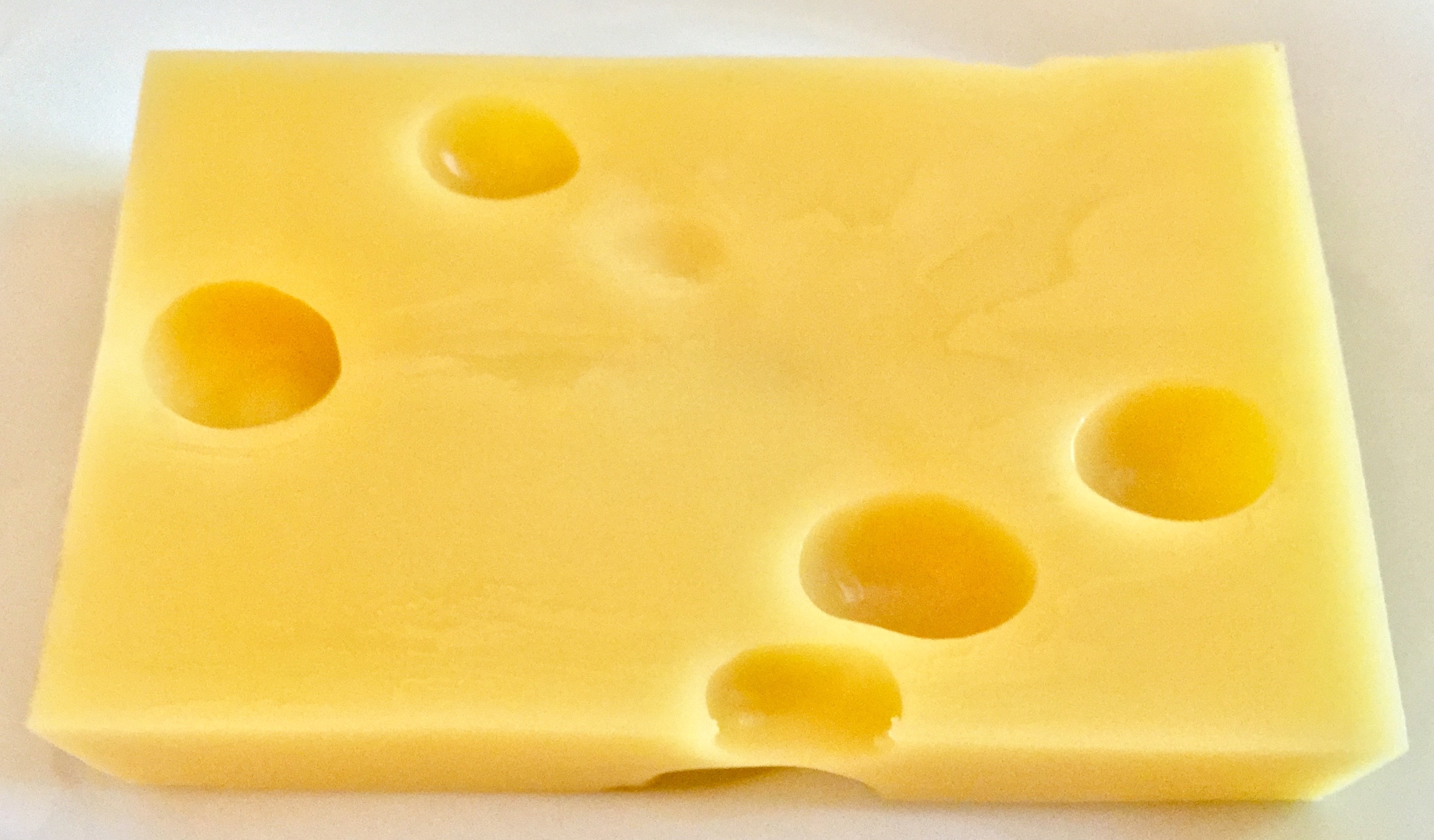}}%
\hspace{0.01\textwidth}
\subcaptionbox{Mozzarella Cheese}{\includegraphics[width=0.312\textwidth]{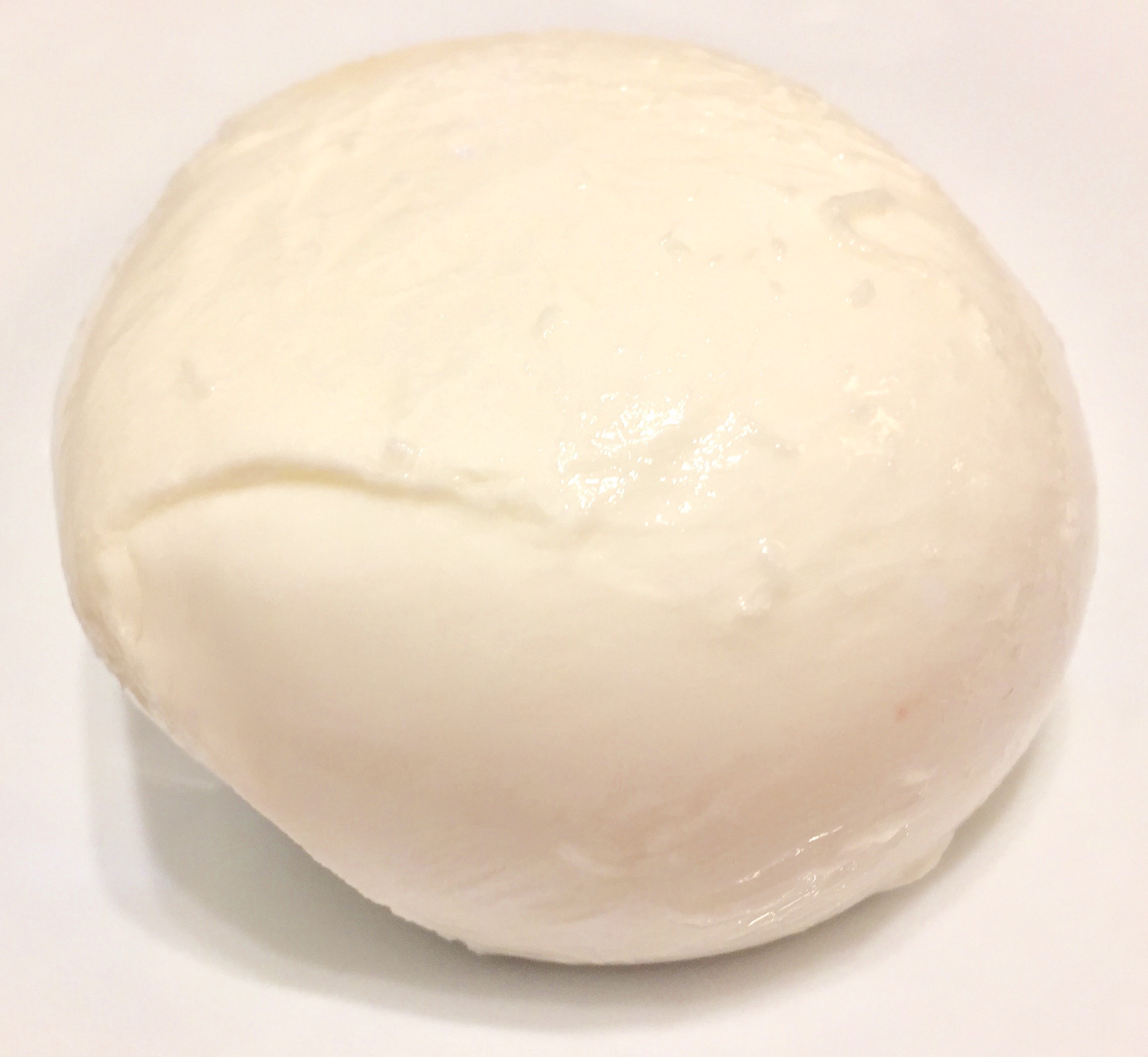}}%
\caption{An example of two topologically-different objects. Homology detects the topological differences by counting the number of holes in the cheeses.}\label{fig:Cheeses1}
\end{figure}

We motivate our use of PH for brain networks by considering different types of cheese and how they differ in their homology.
Calculating homology allows one to differentiate between the shape of a stereotypical Swiss cheese (of the Emmental sort) with holes and the shape of a mozzarella cheese by providing information about the presence or absence of holes in the cheeses. (See Fig.~\ref{fig:Cheeses1} for examples of the aforementioned cheeses.) One can thereby consider the space that surrounds the holes; these are the so-called \emph{loops}. However, homology does not give information about the geometry of the cheeses; for example, it does not `see' that the Swiss cheese is a cube or that the mozzarella cheese is a sphere (unless it happens to be hollow), as it only detects differences in the number of holes.

We now give a brief intuitive introduction to a few concepts behind homology and PH for network data. For more mathematical introductions, see~\cite{Ghrist2008,Edelsbrunner2008,Edelsbrunner2010,otter2017,Stolz2017,Croom}.

\subsubsection{Simplicial complexes} 

To study the characteristics of topological spaces~\cite{Kosniowski1980}, such as the Swiss cheese or the mozzarella cheese, we consider small pieces (`morsels'), on which we can perform computations more easily. When reassembled, the morsels carry the same overall topological information as the original space. We begin building these morsels (i.e., `spaces', to be more formal) using a discrete set of points, which we call `nodes'. We then add `edges' to connect pairs of nodes; `triangles', which consist of three nodes, three edges, and a face; `tetrahedra'; and so on. Formally, these elements are called $k$-simplices, where $k$ indicates the dimension of the simplex. A point is a $0$-simplex, an edge is a $1$-simplex, a triangle is a $2$-simplex, and a tetrahedron is a $3$-simplex.

We can combine different simplices to capture different aspects of a topological space. For example, to capture the holes in the Emmental cheese, we glue together a collection of triangles and edges around the holes; we enclose the same number of holes as in the original cheese. Note that we can only
capture the holes that are enclosed inside the cheese (using the triangles), as one can deform the visible 
holes on the surface into a smooth surface of the cheese. For demonstrative purposes, we therefore assume that the Emmental cheese in Fig.~\ref{fig:Cheeses1} is a cross section of a larger cheese that encloses the holes that are visible in the image. 

\begin{figure}[h!]
\centering
\subcaptionbox{Swiss (Emmental) Cheese}{\includegraphics[width=0.49\textwidth]{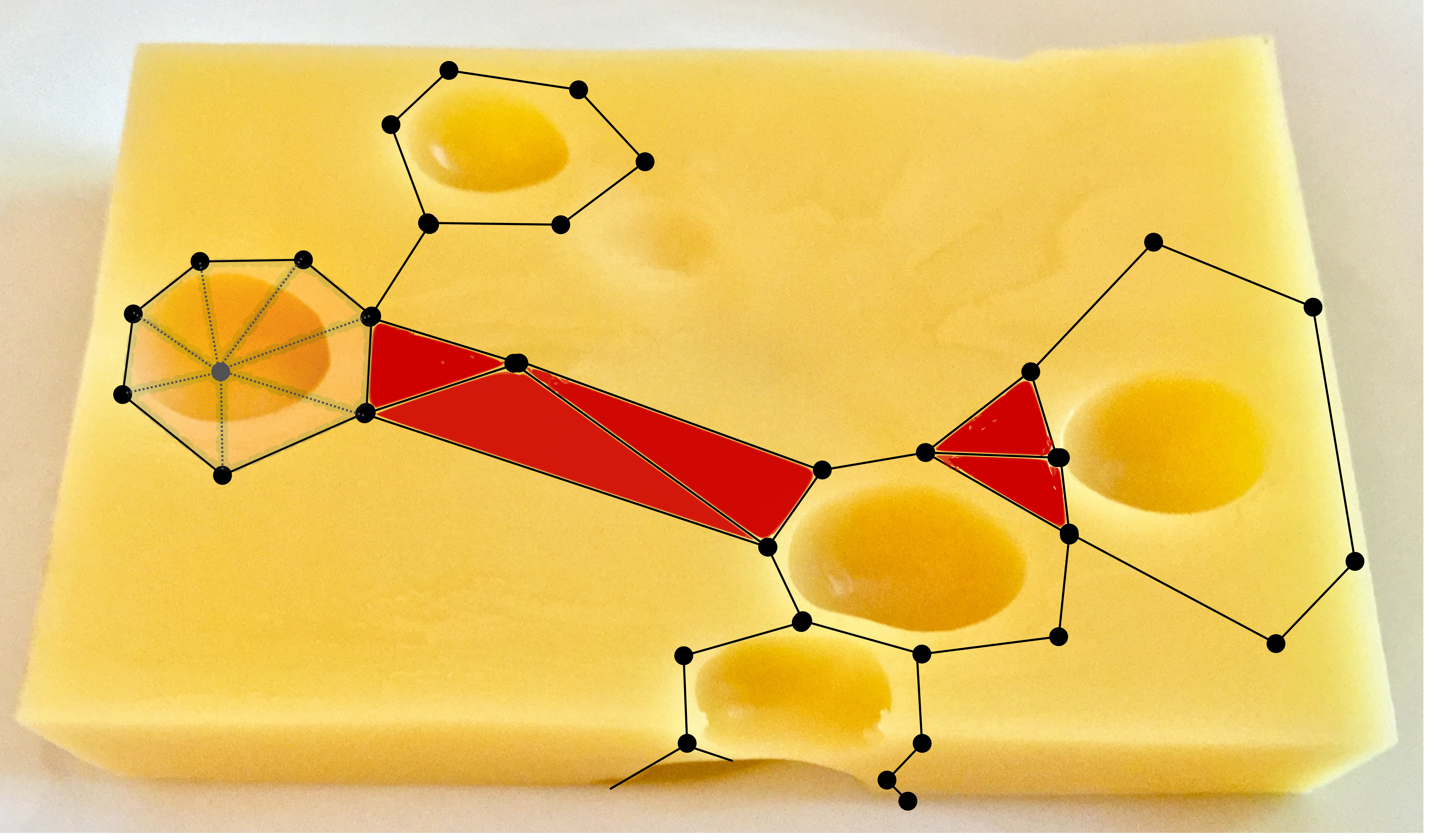}}%
\hspace{0.01\textwidth}
\subcaptionbox{Mozzarella Cheese}{\includegraphics[width=0.312\textwidth]{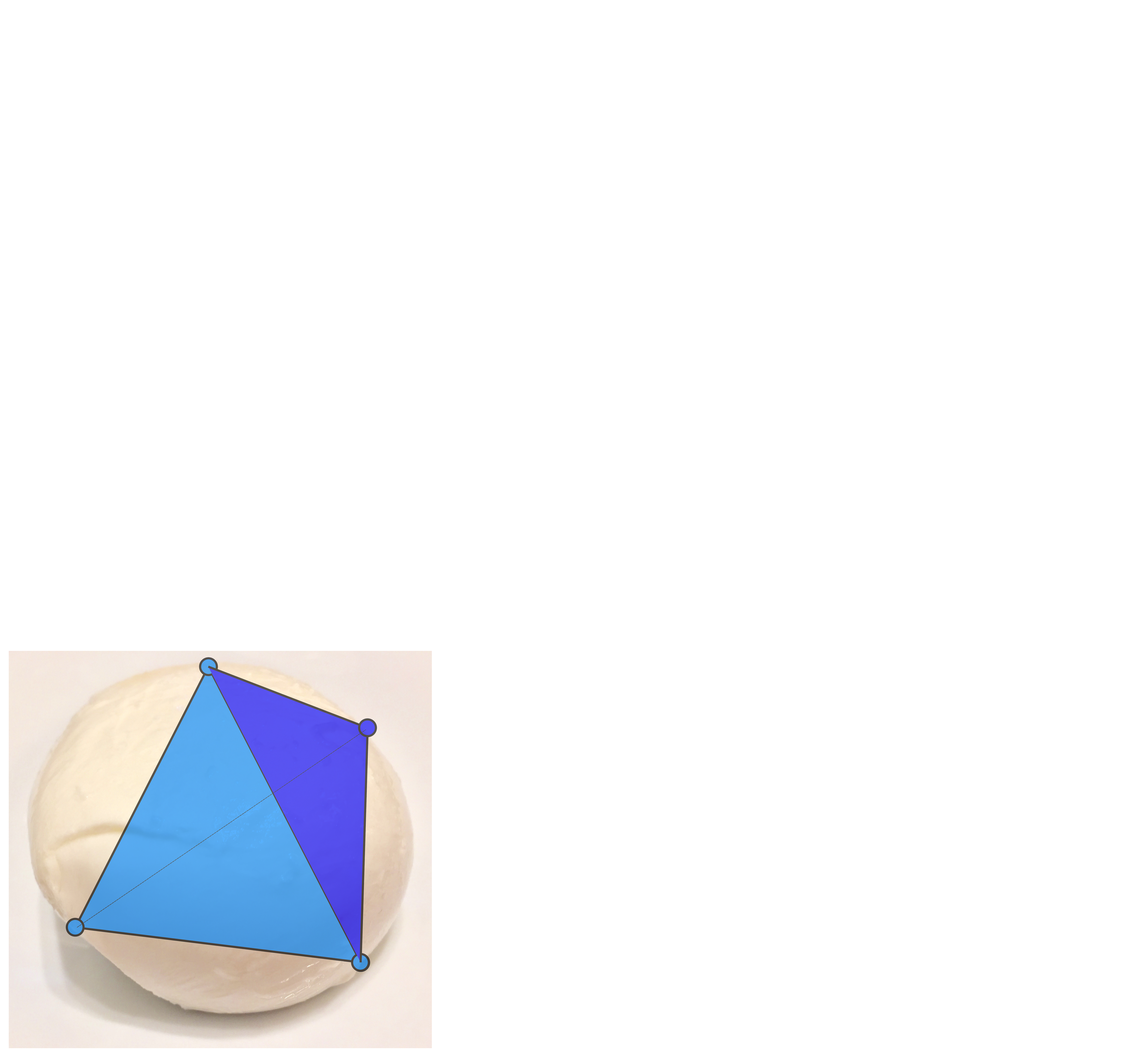}}%
\caption{Simplicial complexes approximate topological spaces and capture their properties. (In the cross section of the Emmental cheese, all holes are topologically the same in the original object. However, we only visualize parts of the simplicial complex in which one of the holes is tiled with triangles but the others are merely surrounded by edges.)
}\label{fig:Cheeses2}
\end{figure}

One can combine simplices to obtain a \emph{simplicial complex} $\Sigma$, and we take the \emph{dimension} of $\Sigma$ to be the dimension of its highest-dimensional simplex. We show examples of simplicial complexes in Fig.~\ref{fig:Cheeses2}, where we again note that we are assuming that the Emmental cheese is only a cross section of a larger hunk of cheese.

\subsubsection{Homology and Betti numbers}

Homology assigns a family of vector spaces (called \emph{homology groups} in more general settings) to a simplicial complex. For a given dimension, the vector spaces capture the topological features in that dimension. For example, for dimension $0$, homology gives a vector space whose elements are connected components; for dimension $1$, homology gives a vector space that has loops as its elements. The dimensions of these vector spaces are called \emph{Betti numbers}, where $\beta_D$ denotes the Betti number for dimension $D$. The first three Betti numbers ($\beta_0$, $\beta_1$, and $\beta_2$) count, respectively, the number of connected components, the number of $1$-dimensional holes (i.e., loops), and the number of $2$-dimensional holes (as found in the Emmental cheese) in a simplicial complex.


\subsubsection{Weight rank clique filtration (WRCF)}

Similarly to being able to distinguish between two types of cheese, we are interested in whether we can use homology 
(and specifically persistent homology) to distinguish between functional networks of schizophrenia patients, siblings of schizophrenia patients, and healthy controls.

In a network, we take a loop to consist of a sequence of four or more nodes and edges that begins and ends at the same node. 
If two loops surround the same hole and can be deformed into one another in the space without tearing open either of the loops, then one counts the loops only once, and we construe them to be different \emph{representatives} (also called \emph{generators}) of a loop. 

To obtain simplicial complexes from a weighted network, we construct a so-called \emph{filtration}. A filtration is a sequence of embedded simplicial complexes that starts with the empty complex:
\begin{equation*}
	\emptyset = \Sigma_0 \subseteq \Sigma_1 \subseteq \Sigma_2 \subseteq \dots \subseteq \Sigma_\text{max} = \Sigma\,.
\end{equation*}
One can obtain a filtration from data in various ways \cite{feng2019}. When given data in the form of a weighted network, the easiest method is to filter by weights \cite{Lee2012}. In the first filtration step, one includes all nodes and the edge(s) with the largest weight in the simplicial complex. In the second step of the filtration, one adds the edge(s) with the second-largest weight to the simplicial complex from step one, and so on. In this way, one obtains a sequence of embedded simplicial complexes that satisfies the properties of a filtration. To construct a \emph{weight rank clique filtration} (WRCF)~\cite{Petri2013}, one performs one additional step: Whenever three edges in a simplicial complex of a filtration form a triangle, one fills in the associated face and one interprets the triangle as a $2$-simplex. Similarly, when four nodes are all connected pairwise by edges, the nodes form a (filled) tetrahedron (i.e., a $3$-simplex). We use the WRCF to analyze our weighted networks. The WRCF has been applied to weighted neuronal networks in several previous studies, including \cite{Petri2013, Petri2014, Giusti2015,Stolz2017}. 

One can use homology to study topological features, such as loops, in every step of a filtration and determine the extent to which a feature persists with respect to the filtration \cite{otter2017}. We say that a topological feature $h$ in a given dimension is \emph{born} at filtration step $m$ if the homology group of $\Sigma_m$ is the first homology group of a simplicial complex in the filtration to include that feature. Similarly, we say that a topological feature \emph{dies} at filtration step $n$ if it is present in the homology group of $\Sigma_{n-1}$ but not in the homology group of $\Sigma_n$. The lifetime of a feature in a filtration is defined as the \emph{persistence} $p$. That is,
\begin{equation} 
	p = n - m \,.
\end{equation}	
If a feature persists until the last filtration step, we say that it has \emph{infinite persistence}. Persistence was first used as a measure to rank topological features based on their lifetime in a filtration in \cite{Edelsbrunner2002}. 

Ideally, one performs a WRCF on a fully connected functional network. However, because of the high computational cost, this is often impossible in practice. We avoid this issue by thresholding our weighted networks before analyzing them.


\subsection{Representations of persistent homology}\label{SubSec:PHRepresentation}

There are multiple ways to represent the output of persistent homology (PH) calculations and to visualize the persistence of topological features and their location within a filtration. The most common representations are barcodes and persistence diagrams. In recent years, a desire to leverage the output of PH computations for machine-learning and data-mining tasks has resulted in the development of alternative representations to both barcode and persistence diagrams \cite{otter2017}. Two of these alternative representations are persistence landscapes~\cite{Bubenik2015I,Bubenik2015} and persistence images~\cite{Adams2015}. In the following subsubsections, we describe barcodes, persistence diagrams, persistence landscapes, and persistence images.


\subsubsection{Barcodes} \label{bar}

\begin{figure}[h!]
\centering\includegraphics[width=\textwidth]{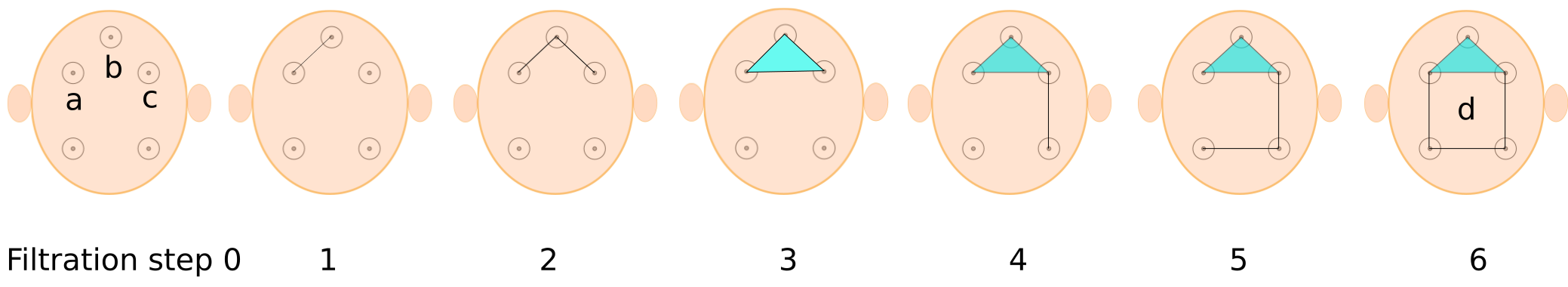}
\centering\includegraphics[width=\textwidth]{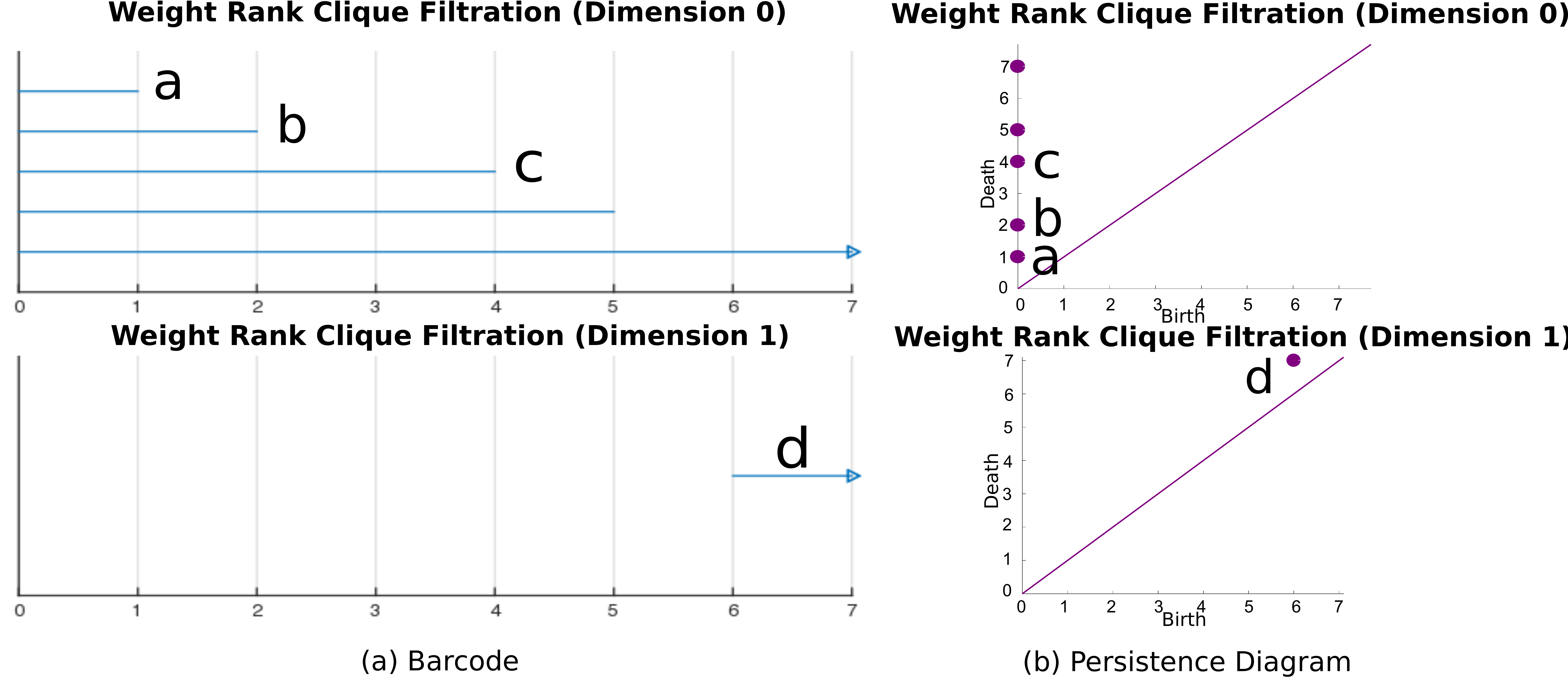}
\caption{Example of a weight rank clique filtration (WRCF) of a neuronal network and the corresponding (a) barcodes and (b) persistence diagrams (PDs) in dimensions $0$ and $1$. The neuronal network consists of different brain regions (indicated by circles), which we interpret as the nodes (indicated by dots) of a network, and weighted edges between the nodes. To construct the filtration, we add the nodes in step $0$, followed by the edge with the largest weight in step $1$, the edge with the second-largest weight in step $2$, and so on. As soon as three nodes are all connected pairwise by edges, we cover the resulting region with a triangle. When four nodes are all connected pairwise, we fill in a tetrahedron. In a $0$-dimensional barcode, we represent each connected component by a bar that starts when the component is born and ends when it dies (e.g., when two components combine with each other). In a $1$-dimensional barcode, each bar represents a loop, which consists of $4$ or more edges and starts and ends at the same node. 
 In a PD, one represents topological features by points, rather than by bars. The distance of a point to the diagonal (the purple line) indicates the persistence of the corresponding feature in the filtration.
}\label{fig:HouseNeuro}
\end{figure}

A common representation of the output of PH calculations is a \emph{barcode}~\cite{Carlsson2005,Ghrist2008}. See Fig.~\ref{fig:HouseNeuro} for an example. A $D$-dimensional barcode is a plot of {a collection of} filtration parameter intervals  $\{[\text{birth},\text{death})_l \}_{l = 1}^t$ that indicate the births and deaths of topological features of dimension $D$. The horizontal axis represents the filtration steps, and each $D$-dimensional topological feature in a filtration is represented by a bar that starts at the filtration step at which the feature is born and ends at the filtration step at which it dies. In a $0$-dimensional barcode, each bar corresponds to a connected component, and the length of a bar indicates how long a particular component is disconnected from other components in a simplicial complex. Similarly, in a $1$-dimensional barcode, each bar corresponds to a loop in a simplicial complex.

\subsubsection{Persistence diagrams} \label{perst}

As an alternative to a barcode, one can use a \emph{persistence diagram} (PD)~\cite{Cohen-Steiner2005}, which is a planar representation of a barcode that conveys the same information. One maps each $[\text{birth}, \text{death})$ interval in a barcode to birth--death coordinates, where the horizontal coordinate of a point represents the birth time of a feature in the associated filtration and its vertical coordinate represents the death time of that feature. Alternatively, one can use a birth--persistence coordinate system, which is particularly useful when examining persistence images (which we will discuss shortly). Points that are farther away from the diagonal identity line represent more-persistent topological features in a filtration. We show an example of a PD in Fig.~\ref{fig:HouseNeuro}. As with barcodes, one can treat PDs as mathematical objects, and one can endow the space of PDs with a distance.


\subsubsection{Persistence landscapes}

A \emph{persistence landscape} (PL)~\cite{Bubenik2015I,Bubenik2015} 
is a sequence of piecewise-linear functions that one can use to visualize and analyze the information in a barcode or PD. Instead of using a bar and its length to represent a feature and its persistence, one now interprets each topological feature as a peak, whose height is determined by the feature's persistence and whose location corresponds to the feature's location in the filtration. In contrast to a barcode or a PD, a PL has three dimensions. As in a barcode, the horizontal axis represents the filtration step. The other two dimensions of a PL are the persistence of a feature and the different layers of the PL. 

\begin{figure}[h!]
\centering\includegraphics[width=\textwidth]{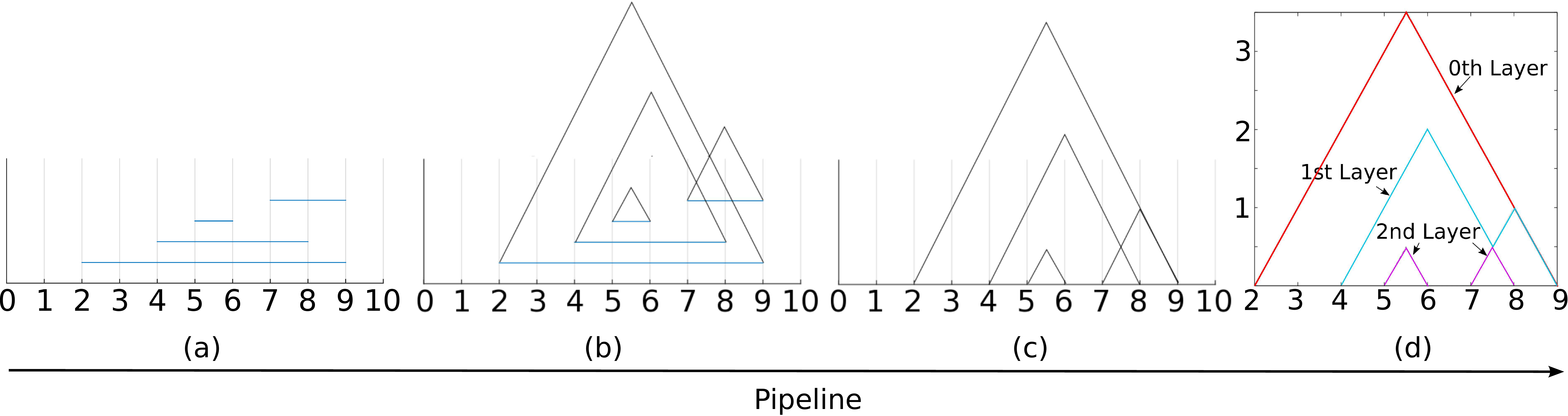}
\caption{Schematic illustration of the steps for converting a barcode into a persistence landscape (PL). We use an example based on a weight rank clique filtration (WRCF) in dimension $1$. (a) Example barcode. (b) One defines peak functions on the bars of a barcode. (c) One collapses the images of the peak functions onto the horizontal axis. (d) The PL consists of the collection of layers $q$ (with $q$ = $0$, $q = 1$, and $q = 2$ in this example), which indicate the $q$th-largest values of the collection of peak-function values. To visualize the third dimension, we show the different layers using different colors.
(This figure is a modified version of a figure in~\cite{Stolz2017}.)
}\label{Fig: Landscapes}
\end{figure}

To create a PL from a barcode, one first defines {a peak function} for each bar. For a given $[\text{birth}, \text{death})$ interval in a barcode, one constructs the function
\begin{equation}
f_{[\text{birth},\text{death})} (x) = \begin{cases} 0 \,, & \text{if } x \notin (\text{birth},\text{death})  \\
		x - \text{birth} 	\,,	&\text{if } x \in \left(\text{birth},\frac{\text{birth}\, + \,\text{death}}{2}\right]  \\
		-x + \text{death} \,,	&\text{if } x \in \left(\frac{\text{birth}\, + \, \text{death}}{2},\text{death} \right) \,.
		\end{cases}
\end{equation}
One then collapses the collection of peak functions onto the horizontal axis of the barcode. 
For a barcode that consists of the collection $\{[\text{birth},\text{death})_l \}_{l = 1}^t$ of intervals, the $q$th layer (with $q \geq 0$) of the PL (i.e., the \emph{$q$}th \emph{PL}) is the following set of functions: 
\begin{align}
	& \lambda_q: \R \rightarrow \R\,, \\
	& \lambda_q(x) = q \text{th-largest value of } \left\{f_{[\text{birth},\text{death})_l}(x) \right\}_{l = 1}^t \,. \nonumber
\end{align}
If the $q$th-largest value does not exist, $\lambda_q(x) = 0$. The $0$th layer of a PL consists of the maximum function values among the collection of functions that one evaluates across a filtration. 
Similarly, the $1$st layer of a PL consists of the second-largest values of the collection of functions that one evaluates across a filtration. One defines other layers  in an analogous way. The \emph{persistence landscape} $\lambda$ of a barcode $\{[\text{birth},\text{death})_l \}_{l = 1}^t$ is defined as the sequence $\{ \lambda_q \}$ of the functions $\lambda_q$. We illustrate the pipeline from a barcode to a PL in Fig.~\ref{Fig: Landscapes}.

An advantage of PLs is that one can construct a mean PL for a set of landscapes. A mean landscape no longer corresponds to a barcode or a PD. However, one can define pairwise distances between two or more mean landscapes and use them to quantify the difference between two sets of barcodes. We use the $L_2$ distance. One can also use a variety of statistical tools on PLs~\cite{Bubenik2015I}. Such calculations have been used for applications like conformational changes in protein binding sites~\cite{Kovacev2015}, the origin of seizures in electroencephalographic (EEG) data from epileptic patients~\cite{Wang2015}, phase separation in binary metal alloys~\cite{Dlotko2016II}, brain geometry in neurodegenerative diseases~\cite{Garg2017}, audio signals in music~\cite{Liu2016}, and motor learning in humans~\cite{Stolz2017}.


\subsubsection{Persistence images} 

Another representation of topological features in PH calculations are \emph{persistence images} (PIs), which are based on PDs and take the form of real-valued vectors \footnote{We use the term `vectorization' for the production of such a vector from a PD. One PD produces one PI, which yields one vector after it is reshaped.} that one can use as an input to a variety of machine-learning approaches. The transformation from a PD to a PI is stable with respect to the $1$-Wasserstein distance and maintains a clear and interpretable connection to the original PD \cite{Adams2015}. For example, PIs have been used to classify different types of neurons~\cite{Kanari2019,Kanari2018}.

\begin{figure}
\begin{centering}
\includegraphics[width=\textwidth]{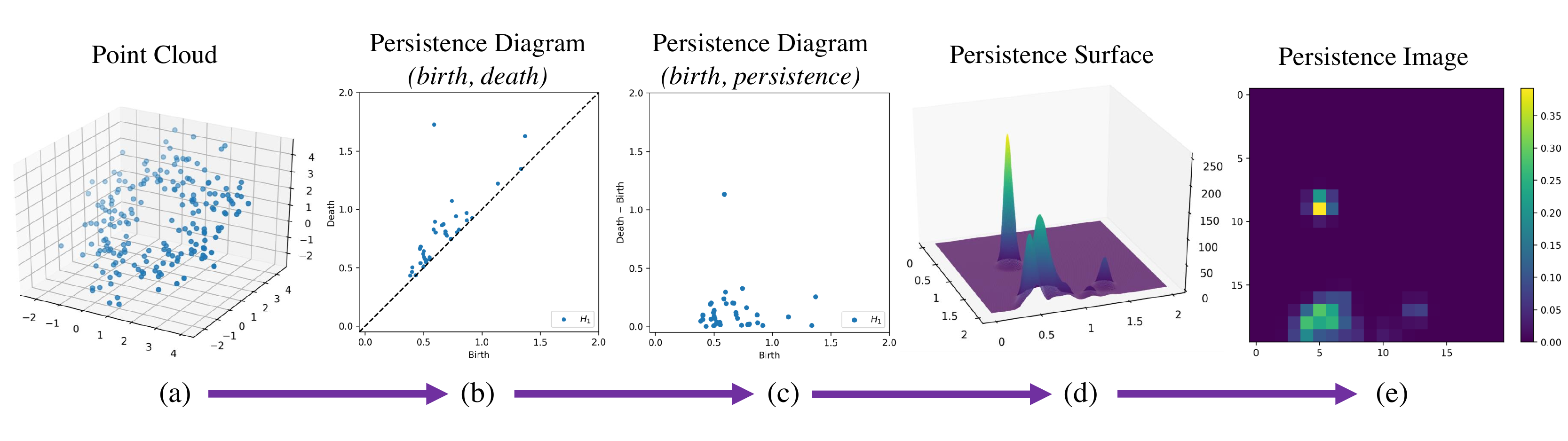}
\caption{Schematic that illustrates the primary steps for converting a persistence diagram (PD) to a persistence image (PI). (a) Sample point cloud in $\mathbb{R}^{3}$. 
(b) PD in birth--death coordinates (i.e., the standard choice), with the diagonal identity line shown. (c) PD in birth--persistence coordinates. (d) The persistence surface generated by centering 2D Gaussian distributions at each point in panel (c). (e) One generates a PI by summing the volume under the 2D Gaussian distributions over the area of a pixel (i.e., the area of a square) in a uniformly-spaced grid overlay. 
}
\label{fig:schematic}
\end{centering}
\end{figure}

We show a schematic of the mapping from a PD to a PI in Fig.~\ref{fig:schematic}, which depicts the various stages that are involved in the transformation. Recall that the output of a PH computation is a set of points (or intervals) corresponding to the birth and death times of each topological feature for a specified homological dimension. In Fig.~\ref{fig:schematic}(b), we show a sample PD \footnote{This example is the output of running a WRCF for dimension $1$ on the functional network of the first sibling from the first time regime in our data set.} for the sample point cloud from Fig.~\ref{fig:schematic}(a). In Fig.~\ref{fig:schematic}(c), we show the PD in the birth--persistence coordinate system. In Fig.~\ref{fig:schematic}(d), we show an overlay of the surface that is generated by centering a two-dimensional (2D) Gaussian on each point in the rotated PD in Fig.~\ref{fig:schematic}(c). Finally, in Fig.~\ref{fig:schematic}(e), we show an example PI that is produced by computing the volume under the surface in Fig.~\ref{fig:schematic}(d) over a uniformly-spaced grid. (We set the resolution so that we have a $20\times20$ grid of elements.) One can then reshape this final PI into a vector by stacking the columns (or, equivalently, the rows), as is often done in image processing. As described in \cite{Adams2015}, the generation of a PI involves the choice of (1) a 2D probability density function to center at each point in the birth--persistence PD, (2) a resolution, and (3) a weighting function.  The role of the weighting function is, when necessary, to 
suppress points in a PD that lie very close to the diagonal and are often construed as `noisy' features. For all of the PIs that we examine in the present paper, we use the default settings for the code: 2D Gaussian probability density functions, a linear weighting function, and a $50\times50$ grid of elements. We choose additional parameters that are associated with these choices (e.g., the variance of the Gaussians) according to the defaults in \cite{PIcode}. When analyzing and comparing multiple PIs, there is an additional pair of values that one must choose based on the data; these are the maximum birth and persistence values. We will see in Section~\ref{subsec: PIs} that this choice can influence our results.

\subsubsection{Software employed}

For our PH calculations, we implement {\sc Matlab} code {that we construct} using {\sc javaPlex} \cite{javaPlex}, a software package for PH \footnote{For an overview of available PH software and additional references, see \cite{otter2017}.}. For a given filtration of a simplicial complex, \textit{{\sc javaPlex}} can output $[\text{birth}, \text{death})$ barcode intervals, representatives for each topological feature, and PDs. It outputs PDs in standard birth--death coordinates, from which one computes birth--persistence coordinates as $(\text{birth},\, \text{death}-\text{birth})$. For the WRCF, we also use a maximal clique-finding algorithm (that is based on the Bron--Kerbosch algorithm \cite{bron1973}) from the Mathworks library \cite{Wildmann2011}. For the analysis and interpretation of our barcodes, we use the {\sc Persistence Landscapes Toolbox}~\cite{Bubenik2015}. We create PIs using the code at~\cite{PIcode} with the default parameters. 


\subsection{Clustering methods from data mining and network analysis}

Given the output of PH calculations, it can be insightful to use clustering methods to compare the PHs of different networks. There are myriad ways to proceed. In the present paper, we use a few different approaches. First, we apply the $k$-means clustering algorithm and community detection to examine whether we can separate the three subject groups based on the topological features of their functional networks. Second, we apply a linear sparse support vector machine (SSVM) to identify pixels in PIs to discriminate between the subject groups and examine which brain regions are generators of loops that help discriminate between groups. We describe these techniques in the following subsubsections.


\subsubsection{Employing $k$-means clustering for subject-group separation}

The method of $k$-means clustering aims to produce a partition of a metric space into $k$ clusters of points \cite{siamcluster}. Suppose that there are $\mu$ data points in a metric space. One selects $k$ of the $\mu$ points as `centers' and assigns all other points of a data set into clusters based on their closest center point. The `score' of such a clustering is the sum of the distances from each point to its nearest center. The desired output of $k$-means clustering is an assignment of points to clusters with the minimum clustering score. However, an exhaustive search for a global minimum is often prohibitively expensive. 
A typical approach to search for a global minimum is to choose a large selection of $k$ initial centers uniformly at random, iteratively improve each selection of centers until the clustering score stabilizes, and then return the identified final clustering with the lowest score for each initialization. One iteratively updates the centers by setting the new center to be the mean of the points that are assigned to the center in the current iteration. One can apply $k$-means clustering either to a distance matrix (which one can calculate for either PDs or PLs) or to a set of input vectors (such as those that one obtains from a PI). 


\subsubsection{Community detection for persistence-landscape classification}\label{Sec: Community Detection} 

Community detection is a method from network analysis that attempts to partition a network into sets (called `communities') of nodes that are more densely connected to themselves than to other sets of networks~\cite{Newman2018,Porter2009,fortunato2016}. One can detect communities in either weighted or unweighted networks. In a weighted network, one finds larger total edge weights within communities than between them.

One can also use community detection to partition data (e.g., for classification) by studying a given distance matrix of data objects (such as mean PLs). We interpret the $n$ PLs as $n$ nodes of a network and convert the pairwise distances between them into edge weights, where a large edge weight signifies closeness in the distance matrix and a small edge weight signifies a long distance between two landscapes. We convert the distance $d(i,j)$ between landscapes $i$ and $j$ into an edge weight $A_{ij}$ between nodes $i$ and $j$ with the following formula:
\begin{equation}
	A_{ij} = 1.01 - \frac{d(i,j)}{\max_{i,j \in\{1, \ldots, n\}}\{ d(i,j)\}}\,.
\end{equation}
This yields an adjacency matrix $A$ with elements $A_{ij}$. Naturally, there are many choices for converting from pairwise distances to pairwise weights, and one has to be careful about how that influences community structure and other network computations.

There are numerous methods that one can use for community detection in networks \cite{fortunato2016}. One approach for decomposing a network into communities (i.e., for performing a `hard partitioning') is to seek a partition that maximizes an objective function $Q$. The objective function that we use is modularity
\begin{equation}\label{eq:QF}
	Q = \sum_{i,j}[A_{ij}-\gamma P_{ij}] \delta(g_i,g_j)\,,
\end{equation}
where $P$ (with elements $P_{ij}$) is a null-model matrix (which specifies the expected edge weight between nodes $i$ and $j$), the resolution parameter $\gamma$ is a factor that determines how much weight one gives to the null model, and $\delta(g_i,g_j) = 1$ if nodes $i$ and $j$ are in the same community (i.e., if $g_i = g_j$) and $\delta(g_i,g_j) = 0$ otherwise \cite{Porter2009,fortunato2016}. 

For our computations, we use the {\sc GenLouvain} package~\cite{MuchaComm,Mucha2010}, which maximizes $Q$ using a variant of the Louvain algorithm \cite{blondel2008} to algorithmically detect communities in our mean PLs. We vary the weighting factor $\gamma$ (which is often called a `resolution parameter') to compare results for different values of $\gamma$.


\subsubsection{Linear sparse support vector machines for discriminatory feature selection}\label{ssvm}

The $1$-norm, regularized, linear support vector machine (i.e., SSVM) classifies data by generating a separating hyperplane between data points that depends on very few input-space features \cite{Bradley1998,Zhu2003,Zhang2010b}. 
A hyperplane is a flat surface that cuts an ambient space into two parts. One can use an SSVM to identify discriminatory features between different groups of data points. One implements linear SSVM feature selection on data points in the form of vectors, so we can use it on our PIs to select `distinguishing pixels' during classification. In a PI, a \emph{distinguishing pixel} is a bounded region in the birth--persistence coordinate system. For 
clarity, we use the term `distinguishing pixel' to signify a region
that is selected by SSVM and a `feature' to refer to a topological feature from a PH computation. 
During the analysis of our results (see Section~\ref{subsec: PIs}),
we aim to match distinguishing pixels to their corresponding features.

We apply a `one-against-all' (OAA) SSVM to dimension-1 PIs from each subject to identify pixels in PIs that can discriminate between the subject groups.
In a one-against-all SSVM, there is one binary SSVM for each class to separate members of that class from the members of all other classes. In our case, this amounts to defining three hyperplanes: one that separates patients from controls and siblings, one that separates siblings from patients and controls, and one that separates controls from patients and siblings. We use a 5-fold cross-validated SSVM. We specify an optimal separating hyperplane by a normal vector, and we use the term `SSVM weights' for the values of the components of the normal vector. We select distinguishing pixels for each classifier by retaining the vector components (which are pixels in this application) with nonzero SSVM weights, ordering the nonzero SSVM weights by decreasing magnitude, and discarding SSVM weights when the ratio of successive SSVM weights drops below a user-specified tolerance. For details, see~\cite{chepushtanova2014}. 

Given a set of distinguishing pixels, we can see for each subject whether the associated functional network have any loops that are born and persist in the corresponding bounded PI region. If there are loops in this region, we can identify a set of brain regions that are representative of that loop in the network. We are thereby able to leverage PIs to obtain (biologically) interpretable information about the involvement of different brain regions in the task (as measured with fMRI) for different subject types.


\section{Results}\label{Sec:Results}

We now present the results of our PH computations to examine loops in functional brain networks. We focus exclusively on topological features in dimension $1$. Additionally, we perform our computations on all four time regimes as part of one data set, rather than separating the data for each time regime. We run our PH computations on four functional networks per subject. From the PH output, we create either PLs or PIs. We then perform our computations either on (i) the full data set of PLs or PIs of 281 subjects and four time regimes (which gives 1124 landscapes or images, respectively, for the data set) or on (ii) the 12 subject-group means of the landscapes or images (from three subject groups with four time regimes each). We indicate which case we are examining in the relevant subsections.

For both PLs and PIs, we find that there seem to be differences in the topological features of the functional networks between subject groups, although we only observe these for PLs when examining means across groups. To illustrate limitations of the methods, we also discuss results in which we were unable to find differences between subject groups.


\subsection{Results of $k$-means clustering on PLs} 

By applying $k$-means clustering to the mean PLs, we are able to separate siblings of schizophrenia patients from
controls and patients. For these calculations, recall that we use all four time regimes in each of the 12 mean landscapes. 

We construct mean PLs from the 1D barcodes (i.e., the barcodes that represent loops in the networks) for each time regime and each subject group. We obtain 12 mean landscapes and exclude infinitely-persisting bars, because all of our landscapes include persistent infinite features and these tend to dominate the first several layers of the landscapes. Other researchers have excluded layers of landscapes (e.g., the first twenty) to filter out `topological noise' \cite{patrangenaru}. Although we threshold our weighted networks prior to analyzing them, this does not necessarily imply that we lose significant information by disregarding the persistent infinite features. Additionally, such features do not necessarily correspond to the most-persistent features in barcodes, as even features that are born in the last filtration steps are infinitely persisting if they do not die during the filtration. In our case, the presence of persistent infinite features prevented us from discriminating between landscapes based on their pairwise landscape distances. When we considered {the number of infinitely persisting} features without the other features, we did not observe any noticeable differences between the three subject groups.

We calculate a pairwise $L_2$ distance matrix of the mean landscapes, and we then perform $k$-means clustering on the distance matrix (which has $12\times 12$ entries). For $k = 3$, we obtain the expected division of the mean landscapes into patients, controls, and siblings. Although the fact that one can separate the three cohorts based on fMRI data is not a new finding --- see, for example,~\cite{Lynall2010,Rubinov2013,Alexander-Bloch2012,Bassett2008,Liu2008,Fornito2012,Singh2016} for patients versus controls and ~\cite{Sepede2010} for patients versus siblings --- the novelty of our calculation is that $k$-means clustering successfully distinguishes between the three different cohorts based on topological information (in the form of loops) in the functional networks.

\begin{figure}[h!]
\centering\includegraphics[width=\textwidth]{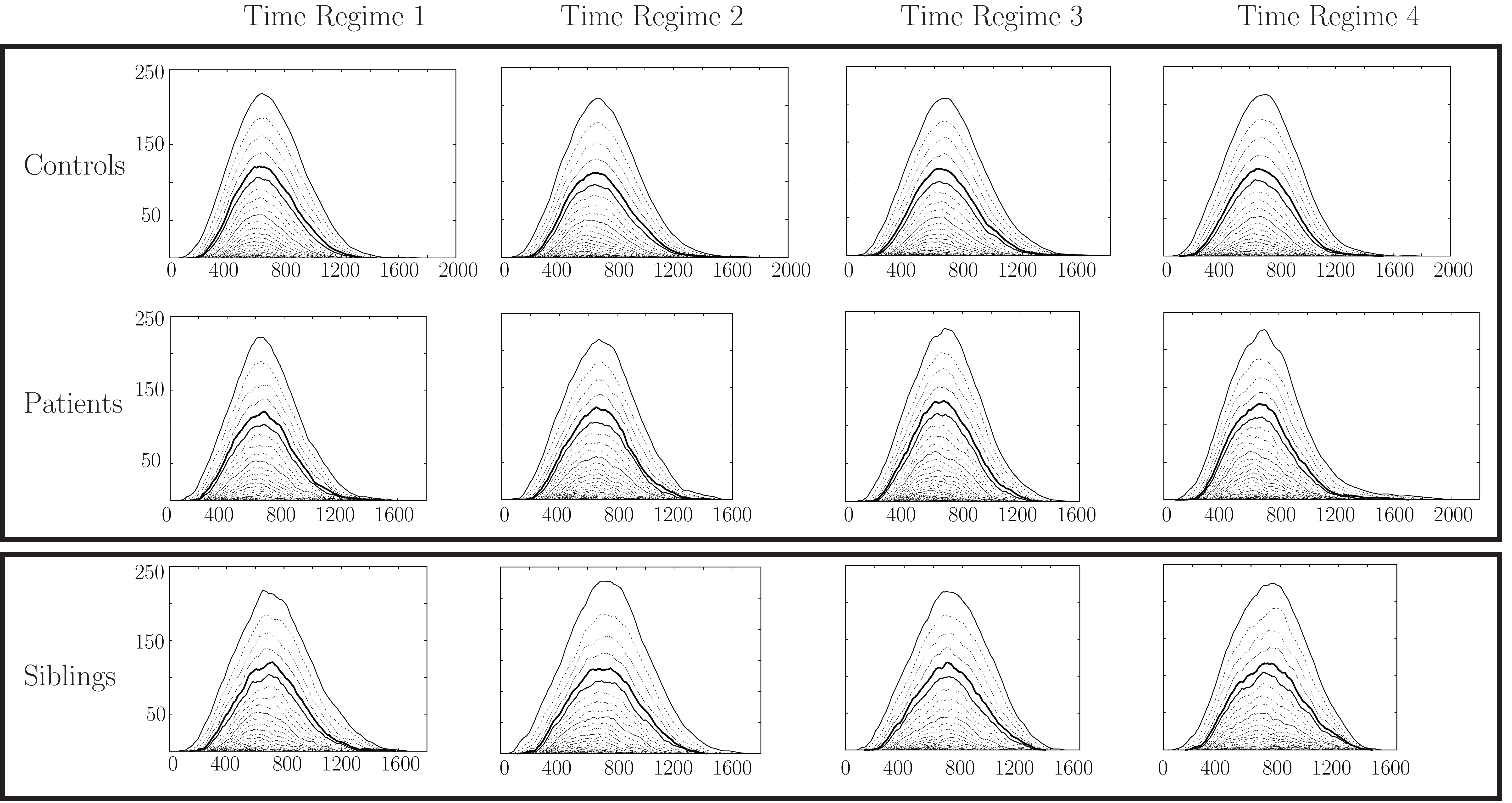}
\caption{Mean PLs for each of the four time regimes and subject groups. Using $k$-means clustering with $k = 2$ on the set of 12 PLs (which consists of all subject-group means and time regimes as one data set) assigns patients and controls to one group. We show the mean PLs and their $k$-means-clustering grouping for the four time regimes separately.
}\label{Fig:k-means Average Landscapes}
\end{figure}

We also perform $k$-means clustering for $k = 2$. Surprisingly, we find that the patients and controls are grouped in one cluster for all time regimes, whereas the siblings are in a separate cluster for all time regimes. We show the mean landscapes and clusters in Fig.~\ref{Fig:k-means Average Landscapes}.

For $k \geq 4$, we do not observe a clear subject-group separation. To compare our results with ones from other clustering methods, we also apply average linkage clustering to the distance matrix and perform community detection on networks that we construct from those distance matrices (see Section~\ref{Sec: Community Detection}). We obtain the same qualitative result for these two methods as we did for $k$-means clustering. For community detection, we observe a clear separation for resolution-parameter values of $\gamma = 0.82, 0.83, \dots, 1.14$ into two communities (the siblings versus the patients and controls). These results appear to support our prior observation that the sibling cohort is particularly distinct from the other two cohorts, as compared to any other pairwise comparison among the three cohorts, with respect to their loop topology in the functional networks.

We also perform a permutation test on the mean PLs for each time regime {to determine the significance of the landscape distances, as suggested in~\cite{Bubenik2015}}. In this permutation test, we regroup the individual landscapes into three groups uniformly at random, create a new mean landscape for each newly assigned group, and calculate the pairwise $L_2$ distances between them. We then count how many of the $L_2$ distances of the new groups are larger than the ones that we observe when using the mean landscapes of the three subject groups. We use 10000 permutations to obtain our results, which we summarize in Table~\ref{Tab:pValuesLandscapes}.

\begin{table}[!ht]
\caption{Using a permutation test, we calculate p-values for the mean landscape distances between the three subject groups in each time regime.}\label{Tab:pValuesLandscapes}
\centering
\begin{tabular}{cccc}
\br
p-values for &Controls versus Patients & Controls versus Siblings & Patients versus Siblings \\ \mr
time regime 1 & 0.302 & 0.200 & 0.051 \\ 
time regime 2 & 0.460 & 0.009 & 0.052 \\ 
time regime 3 & 0.477 & 0.102 & 0.270\\ 
time regime 4 & 0.736 & 0.110 & 0.229\\ \br
\end{tabular}
\end{table}

Interestingly, for time regimes $1$ and $2$, we find {almost} significant distances (i.e., the p-values are slightly larger than $0.05$) between the patient and sibling mean landscapes, whereas the p-values for time regime $3$ and $4$ suggest that the distance is not significant (even though the p-values are {small in comparison to the p-values that we observe for controls versus patients}). The distance between the mean landscapes of the controls and the siblings appears to be significant for time regime $2$, but this does not appear to be the case for the other time regimes, although the p-values are again much smaller than for the distances between the mean landscapes of the patients and controls. For the controls and the patients, there are many other divisions into two groups that lead to more extreme distances between the mean landscapes than what one obtains by simply assigning them to a control group and a patient group. 

To see if we can further support our result from $k$-means clustering for $k = 2$, we artificially group the controls and patients into one group to create a mean landscape and again perform a permutation test to verify whether the distance between the mean landscapes for the two groups is significant. In Table~\ref{Tab:pValuesLandscapes2}, we show the p-values that we obtain with 10000 permutations.

\begin{table}[!ht]
\caption{Using a permutation test, we calculate p-values for the controls-and-patients mean landscape versus the siblings mean landscape.}\label{Tab:pValuesLandscapes2}
\centering
\begin{tabular}{cccc}
\br
Time regime 1 &Time regime 2 & Time regime 3 & Time regime 4 \\ \mr
0.112 & 0.008 & 0.092 & 0.110 \\ \br
\end{tabular}
\end{table}

For time regime 2, we obtain a significant distance, but the p-values for time regimes 1, 3, and 4 are about $0.1$. Given the artificial grouping of the two subject groups, we interpret these values as small, although they are not statistically significant.


\subsection{Results of community detection using a distance matrix from individual PLs}

We construct PLs from each of the 1D barcodes, which we calculate by examining each subject in each of the four time regimes, and we calculate the $L_2$ distance matrix for the resulting 1124 PLs. We again use the distance matrix to construct a network between the PLs, and we detect communities in this network by maximizing modularity. For $\gamma = 0.92, 0.93, \dots, 1$, we obtain a separation into two communities. The partition that is closest to what we observe with $2$-means clustering for the mean landscape distance occurs for the resolution-parameter value $\gamma = 0.93$. We summarize our results in Table~\ref{Tab:CommunityClusters}.

\begin{table}[!ht]
\caption{Number of subjects from each subject group that are assigned to communities $1$ and $2$ by community detection using modularity maximization.}\label{Tab:CommunityClusters}
\centering
\begin{tabular}{ccc}
\br
Subject group & Number of subjects in community $1$ & Number of subjects in community $2$ \\ \mr
Patients & 122 & 94 \\
Controls & 418 & 290 \\ 
Siblings & 93 & 107\\ \br
\end{tabular}
\end{table}

We also apply $k$-means clustering and average linkage clustering to the distance matrix from the individual PLs (results not shown).
Of all of the classification methods that we perform on these distance matrices, community detection appears to perform best at `separating' the subject groups, although we do not observe a very clear separation. 


\subsection{Results from our analysis of PIs}\label{subsec: PIs} 

We find that PIs can identify discriminatory topological features across the three subject groups that we considered.
We generate PIs for each of the subjects for each of the four time regimes for the 1D PDs. We set the resolution, probability density function, and weighting function to the defaults in the PI code at \cite{PIcode}. 

\begin{figure}[h!]
\centering
\subcaptionbox{}{\includegraphics[width=0.49\textwidth]{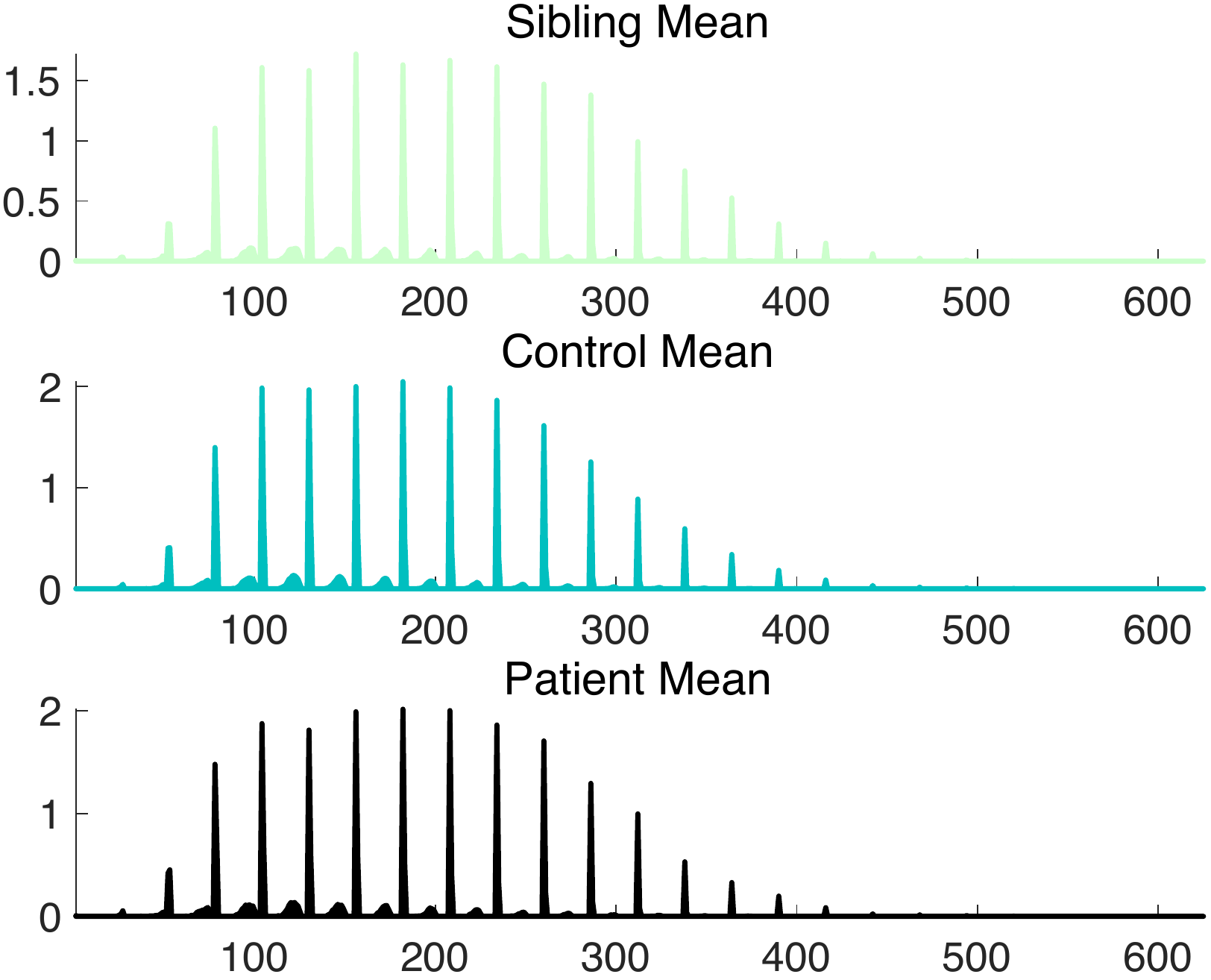}}%
\hspace{0.01\textwidth}
\subcaptionbox{}{\includegraphics[width=0.49\textwidth]{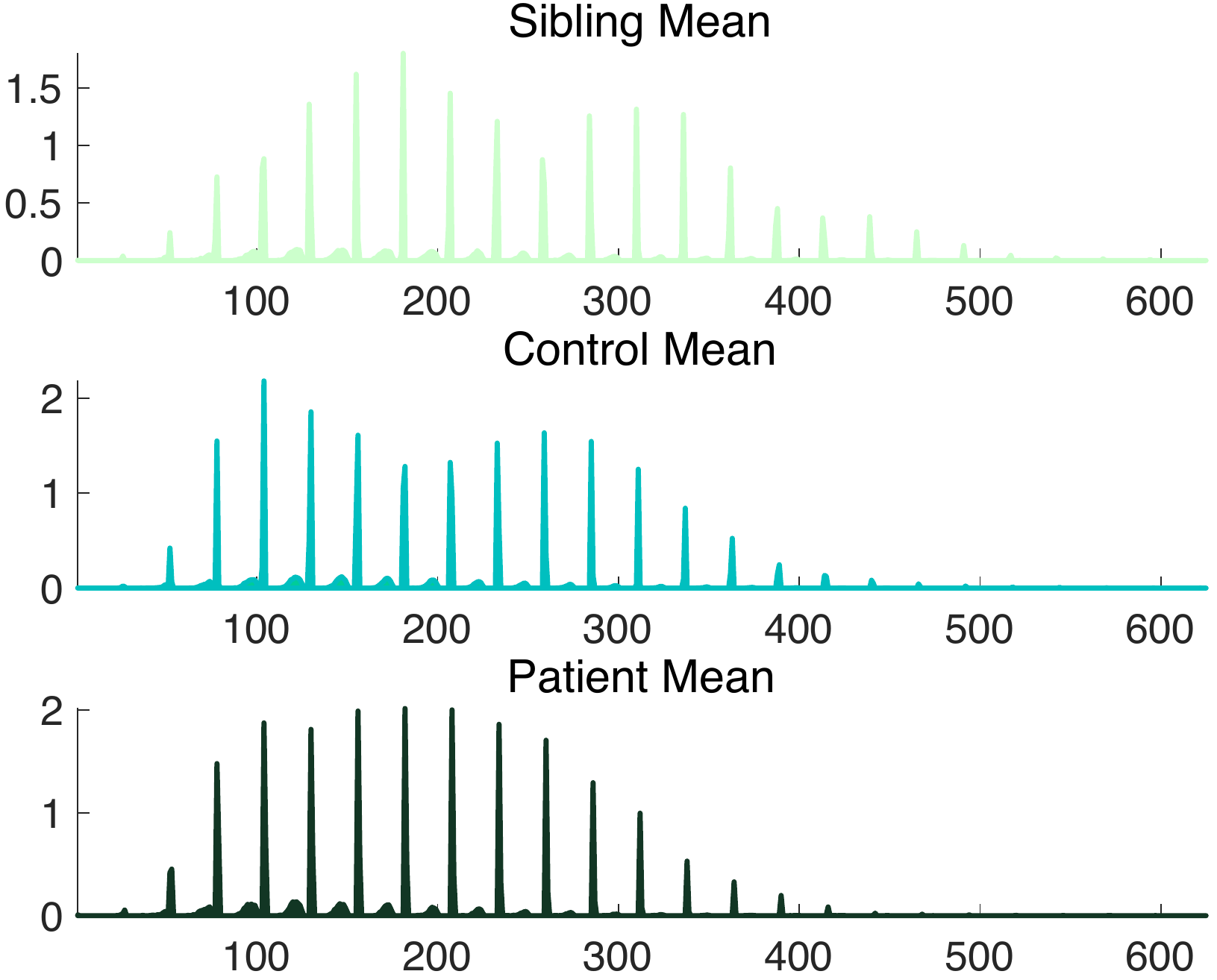}}%
\caption{(a) The mean vectorized PI for each subject group that we generate using the maximum values of birth and persistence across all subjects to create the PIs. We then take the means over the PIs of each group. The horizontal axis corresponds to individual pixels in the PIs, and the vertical axis indicates their intensity values. (b) Mean vectorized PI for each subject group that we generate using maximum values of birth and persistence that we determine by calculating the maximum birth and persistence for each of the three groups separately and using this group-specific information to create the PIs for each subject within its group. We then take the means over the PIs of each group.
}\label{fig:PIstatsSame}
\end{figure}

\begin{figure}
\begin{centering}
\includegraphics[width=.6\textwidth]{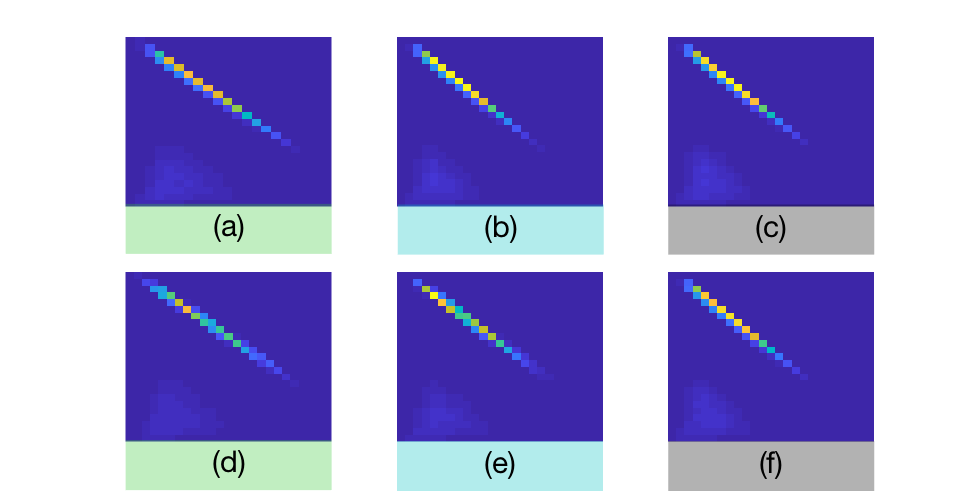}
\caption{The mean PI for each subject group. We generate panels (a)--(c) using the maximum values of birth and persistence across all subjects to compute all PIs before creating the depicted means over the PIs in each subject group. We generate panels (d)--(f) using the maximum values of birth and persistence using the maximum birth and persistence separately for each subject group to compute PIs within each group before creating the depicted means over the PIs in each group. The color axis is the same across rows. From left to right in each row, we show the mean PIs for siblings, controls, and patients.}
\label{fig:PIstatsImages}
\end{centering}
\end{figure}

There is an additional pair of values ---  the maximum birth and maximum persistence values --- that one must choose from the data that one is analyzing using PIs. These values determine the discretization of the pixel boundaries in the PIs once one sets the resolution. Possibilities include taking the maximum birth and persistence values across all PDs or normalizing each PD relative to its individual maximum. In the original paper on PIs~\cite{Adams2015} took maximum values across all PDs under consideration, although no theoretical rationale was provided for this choice. We were unable to obtain clear results using either of the above two conventions. For example, in the left image of Fig.~\ref{fig:PIstatsSame}, we see the mean vectorized PI for each subject group when we generate the PIs using the maximum birth time and maximum persistence across all subjects. (We create the mean vectorized PI for each subject group by taking the mean of each vector entry. {We take the mean across all four time regimes to ensure that we have enough data to make a meaningful comparison.}) Observe that the means look very similar, aside from a slight variation in their amplitudes. For each group, the mean PI is the mean of the vectorized PIs for all of the individuals in the group. The top row of Fig.~\ref{fig:PIstatsImages} has the mean PIs in image form when one selects the maximum birth and persistence across all subjects.

Alternatively, if we use a priori knowledge of subject-group membership and fix the maximum birth values separately for each subject group (based on the collection of PDs that we compute separately for each subject group), we can discriminate between the three subject groups. This provides a first interesting observation from the PIs: there appears to be nontrivial information in the maximum birth time that corresponds --- or almost corresponds, in exceptional cases in which multiple edges have exactly the same weight --- to the number of pairs of regions in the brain with positive functional connectivity. (Recall that we do not add edges that correspond to negative Pearson correlations.) In Fig.~\ref{fig:PIstatsSame}(b), we show the mean vectorized PI for each subject group, where we set the maximum birth and maximum persistence values separately for each subject group (instead of setting the maximum birth and persistence values to be the same for all subjects). Observe that the sibling and control means both have two humps, whereas the patients have one hump that is clearly discernible. 
Similarly, in Fig.~\ref{fig:PIstatsImages}, we observe two patches along the prominent diagonal with high intensity for the means of the siblings and the controls. However, in the bottom row, we only observe one clear (and elongated) hot spot for the mean of the patients. Therefore, there are multiple, smaller regions in which loops often occur in the filtrations of the functional networks of siblings and controls, whereas there is seemingly a single, larger region of loops in the filtrations of the networks of the patients. 

\begin{figure}[h!]
\centering
\subcaptionbox{}{\includegraphics[width=0.49\textwidth]{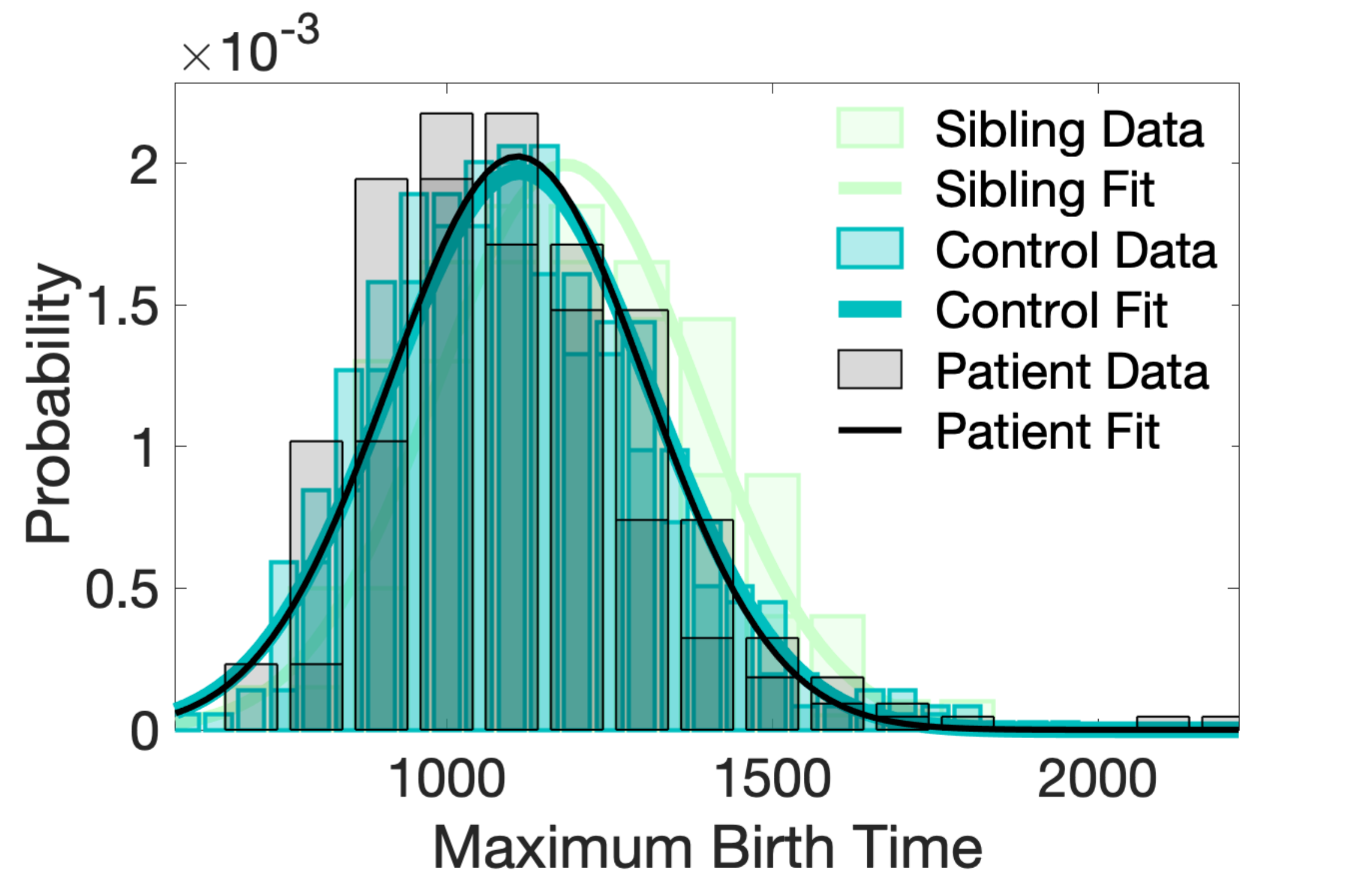}}%
\hspace{0.01\textwidth}
\subcaptionbox{}{\includegraphics[width=0.49\textwidth]{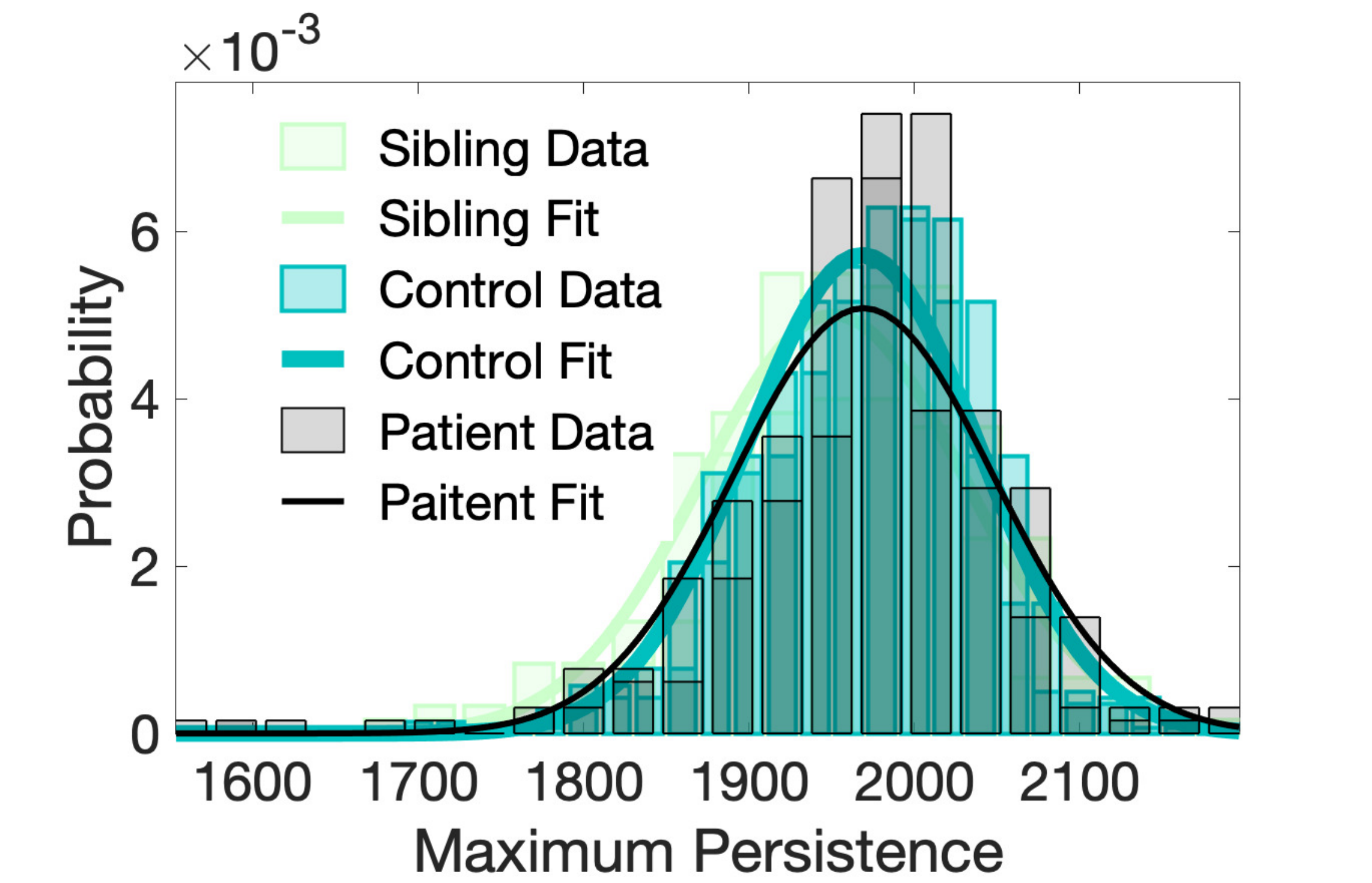}}%
\caption{The distribution of (a) the maximum birth times across all samples for each subject type and (b) the maximum persistences across all samples for each subject type. 
}\label{fig:dists}
\end{figure}

It is also worth commenting on the locations of the local maxima for each subject type. Relative to the maximum values across each class, groupings of loops occur at different locations. From the values of the vectors, we see that the controls and patients have more similar maximum magnitudes than do the patients and their siblings. Therefore, we conclude that we are able to accurately separate the populations using PIs. Surprisingly, despite the pronounced difference in SSVM performance based on the PIs when we use different maximum values for each class, the distributions of the maximum birth times and maximum persistences for each subject type are not statistically-significantly different from each other. In Fig.~\ref{fig:dists}, we show Gaussian fits to the set of maximum birth times and maximum persistences for each subject type. Observe the strong similarity across all classes and the especially close similarity between the control and patient distributions. Because the maximum values are linked closely to the preprocessing of the data, it is important to conduct further research into how to account for these observations. One can also normalize PIs in other ways, such as by normalizing each PI individually by its own maximum value. However, we find that such individual normalization of PIs obscures information (specifically, the maximum birth time of each subject group) that appears to be relevant. The following results are based on the PIs that we generated using the maximum values that we determined by class membership.

\begin{figure}
\begin{centering}
\includegraphics[width=\textwidth]{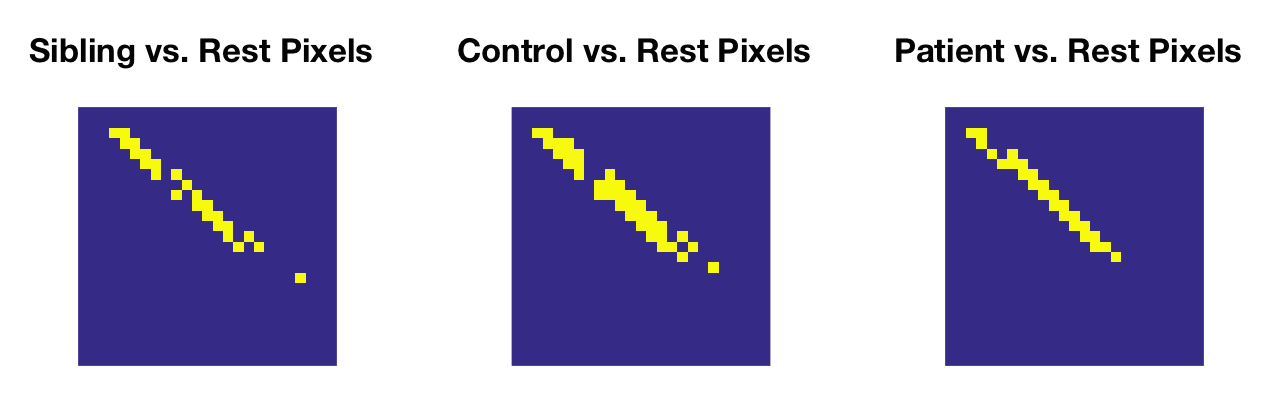}
\caption{The set of 41 distinguishing pixels (of 625 in total) that we determine via SSVM to be critical for obtaining 100\% classification accuracy on the testing set of PIs.
}
\label{fig:selectedfeatures}
\end{centering}
\end{figure}

As we discussed in Section~\ref{ssvm}, it is possible to apply a linear SSVM to the set of PIs to identify distinguishing pixels to help us interpret our classification results. Using a one-against-all SSVM with 5-fold cross validation, we obtain a 100\% classification accuracy. In Fig.~\ref{fig:selectedfeatures}, we show the distinguishing pixels from each of the three binary classifiers. By taking the union of these pixels, we obtain $41$ distinct pixels from the total of $625$ pixels in the PIs. We again emphasize that each of these pixels corresponds to a bounded region in the birth--persistence plane.

Interpreting these distinguishing pixels requires discussing their relationships with particular regions of the brain. We make these connections as follows. For each subject, it is possible to determine whether a topological feature (in our case, a loop) in a filtration of a network exists in the bounded region of the birth--persistence plane that corresponds to a particular distinguishing pixel. If a loop does exist, one can identify a set of brain regions (see Section~\ref{Sec:Discussion}) that comprise the loop (i.e., representatives of this loop). We are particularly interested in brain regions that are consistently involved in the generation of particular loops across subjects. We identify the set of nodes, which we call \emph{top node(s)} \footnote{One can interpret our calculation of top nodes in a similar spirit as calculations of node centralities \cite{Newman2018}.}, that are involved in the generation of loop(s) for each distinguishing pixel in each of the four time regimes for each subject. We then create histograms of the union of the nodes that we select in this fashion to examine the relative importances of top nodes across each subject type. 

\begin{figure}[ht!]
\centering
\subcaptionbox{}{\includegraphics[width=0.49\textwidth]{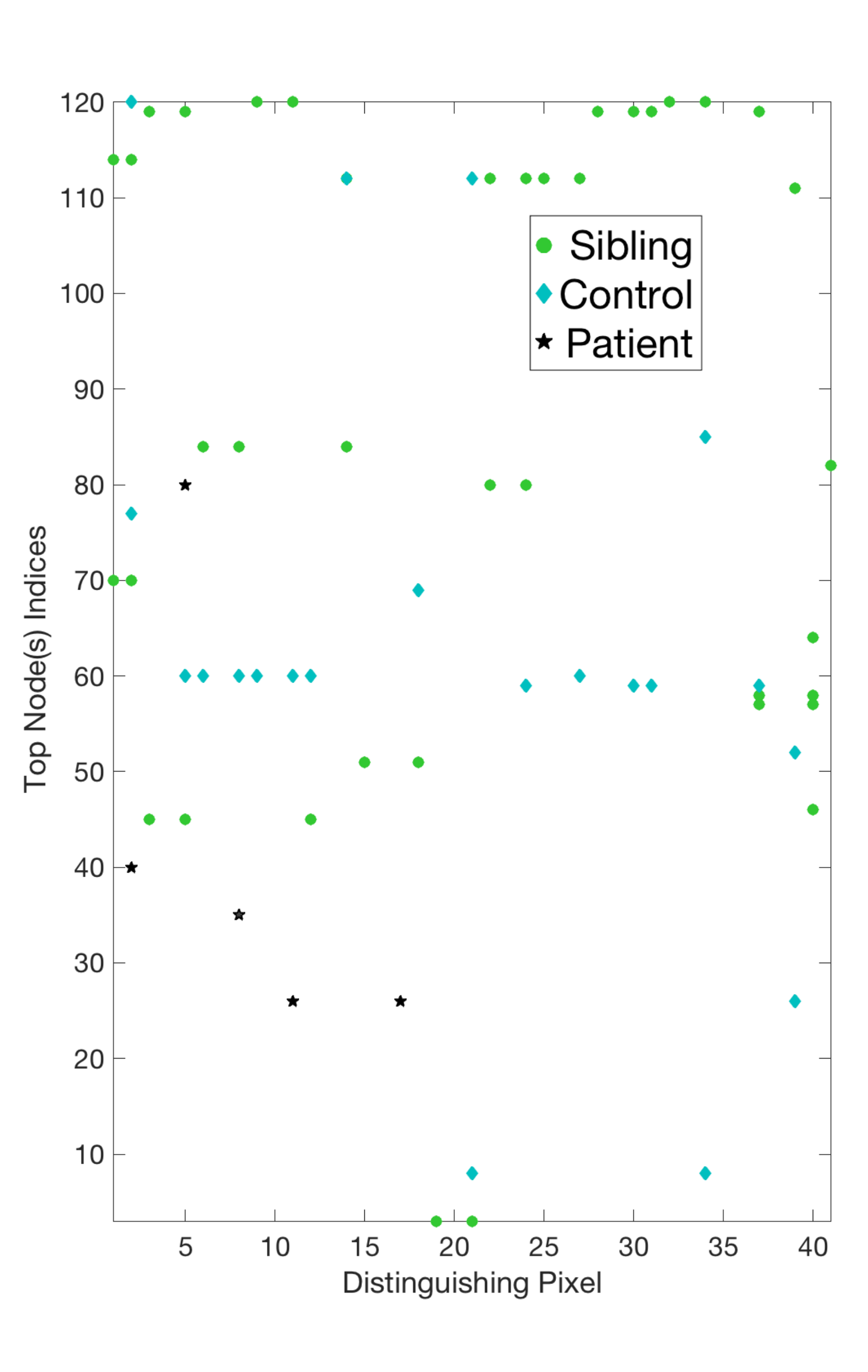}}%
\hspace{0.01\textwidth}
\subcaptionbox{}{\includegraphics[width=0.49\textwidth]{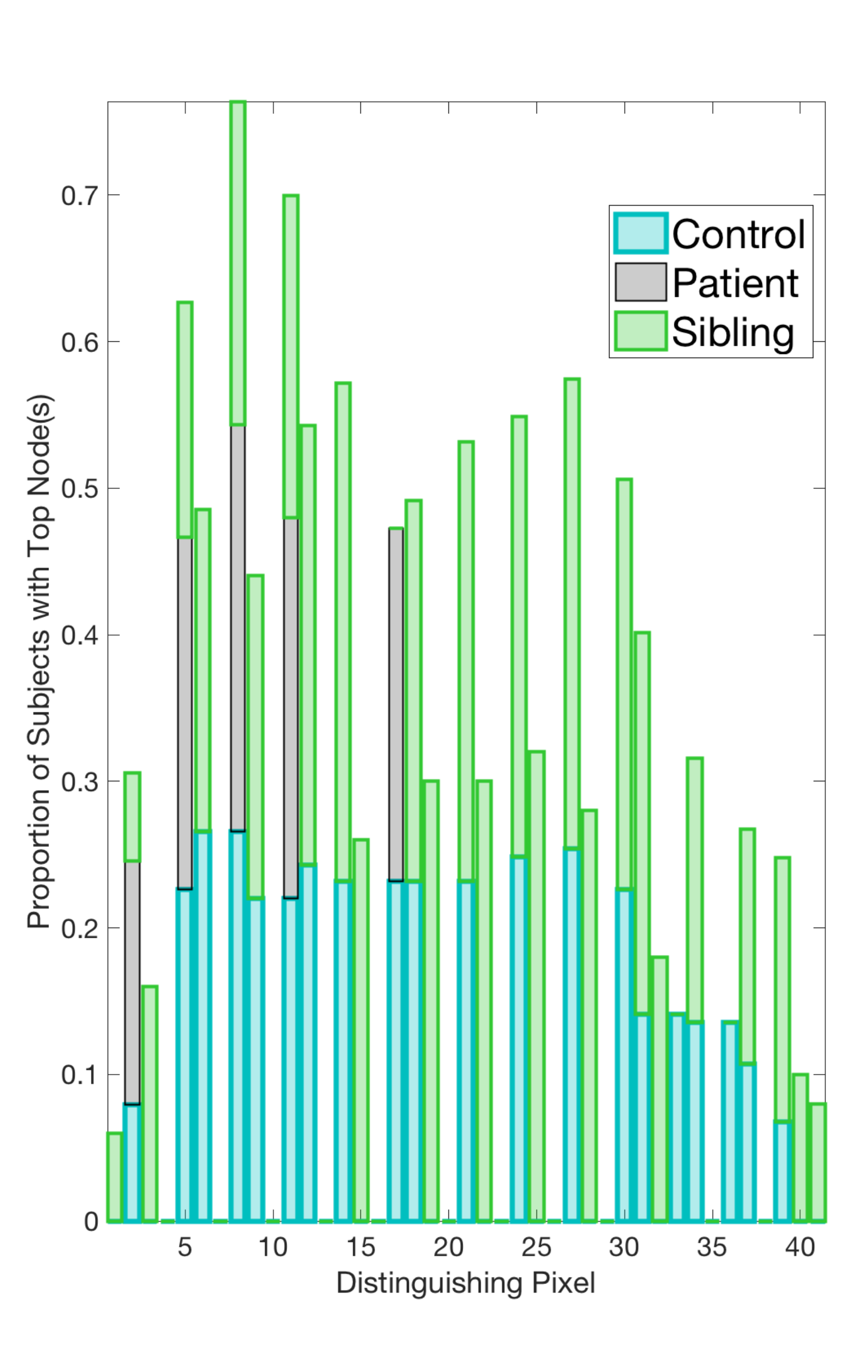}}%
\caption{(a) The index (indices) of the top node(s) that are associated with each distinguishing pixel that we determine via SSVM. (b) A stacked bar graph of the proportion of each subject type that have the corresponding node(s). Pale green indicates siblings, greenish blue indicates controls, and black indicates patients.
}\label{fig:MAX_nodes}
\end{figure}

In Fig.~\ref{fig:MAX_nodes}, we give the relative importances of different brain regions for each pixel. In Fig.~\ref{fig:MAX_nodes}(a), we show the top nodes for each subject type based the proportion of the subject types for which that top node is involved in the generation of a loop in the distinguishing-pixel region. In Fig.~\ref{fig:MAX_nodes}(b), we show the proportion of the subjects for which the top node(s) is (are) present. The vertical gaps in each plot signify that there are no nodes that are consistently involved in loops for that distinguishing pixel. We make several observations from Fig.~\ref{fig:MAX_nodes}(a). First, there are only five distinguishing pixels for which we find top nodes for the patients. Therefore, we are unable to predict which brain regions are involved in loops in the functional networks during the given task for schizophrenia patients. By contrast, there are many distinguishing pixels for which we find top nodes for the siblings. The control group lies between the other two in terms of its number of distinguishing pixels with top nodes, but there are still few top nodes in comparison to the number of distinguishing pixels that have top nodes. In Tables \ref{table:features}--\ref{table:features4} (see also Figs.~\ref{fig:MAX_nodes_Sib}--\ref{fig:MAX_nodes_Pat} of~\ref{S1}), we indicate which brain regions (as well as their locations) we identify as top nodes. We include only the distinguishing pixels in which top nodes exist within a cohort. 

An equivalent way to identify a top node is to calculate the percentage of a given subject class that has a topological feature in the corresponding pixel region (see the bar graph in Fig.~\ref{fig:MAX_nodes}) and determine if a specific node is in the group of representatives for all of the subjects that have a topological feature in the pixel region. We identify a node as a top node if it occurs in the list of representatives for a topological feature for every subject of the class with a topological feature in the pixel region. Therefore, when considering Tables \ref{table:features}--\ref{table:features4}, it is possible for the same brain regions to be listed for more than one distinguishing pixel index. This is also reflected in Fig.~\ref{fig:MAX_nodes} by the occurrence of multiple markers along the same horizontal line.

 \begin{table}
 \centering
 \caption{{\bf Top nodes that are involved in loop representatives within the bounds of distinguishing pixel birth--persistence regions (part I).}
We include only distinguishing pixels for which there is (are) top node(s) within a cohort. `Left' and `Right' refer to the hemispheres of the brain. We use the following abbreviations: superior frontal gyrus medial segment (MSFG), superior temporal gyrus (STG), opercular part of the inferior frontal gyrus (OpIFG), transverse temporal gyrus (TTG), frontal operculum (FO), gyrus rectus (GRe), middle frontal gyrus (MFG), orbital part of the inferior frontal gyrus (OrIFG), precuneus (PCu), cuneus (CC), anterior insula (AIns), superior parietal lobule (SPL), lingual gyrus (LiG), cerebellum exterior (CE), parahippocampal gyrus (PHG), medial frontal cortex (MFC), medial orbital gyrus (MOrG), and posterior cingulate gyrus (PCgG). 
 }
 \label{table:features}
 \begin{tabular}{*7c}
 \br
 Pixel Indices &  \multicolumn{2}{c}{Siblings} & \multicolumn{2}{c}{Controls} & \multicolumn{2}{c}{Patients}\\
 \mr
 {} & Node& Location & Node & Location & Node & Location\\ \mr
1 		& 70 		& Left MSFG		& --	& 	--				& 	--	& 	--	\\
 		& 114		& Left STG		&--	&	--				&	--	&	--	\\
 		\mr
2 		& 	70		& Left MSFG		&	77	&	Left OpIFG	&	40	&	Left FO	\\	 		
		& 114		& Left STG		&120	&	Left TTG		&	--	&	--			\\	 
		\mr
3		& 45		& Right GRe		&--	&	--				&	--	&	--	\\	 
		& 119		& Right TTG		&	--	&	--				&	--	&	--	\\	 	
		\mr
5		& 45	& 	Right GRe			&	60	& Left MFG		&80	& Left OrIFG	\\	
		& 119	& 	Right TTG		&	--	&	--				&	--	&	--	\\	
		\mr 
6		& 	84	& Left PCu			&	60	&Left MFG		&	--	&	--	\\	 
		\mr
8		& 	84	& Left PCu			&	60	&Left MFG		&35 	&	Right CC	\\ 
		\mr
9		& 120	& Left TTG			&	60	&Left MFG		&	--	&	--	\\	 
		\mr
11		& 120	& Left TTG			&	60	&Left MFG		&	26	&	Left AIns	\\
		\mr
12		& 	45	&Right GRe	 		&	60	&Left MFG		&	--	&--		\\	
 \br
 \end{tabular}
 \end{table}

  \begin{table}
 \centering
 \caption{{\bf Top nodes that are involved in loop representatives within the bounds of distinguishing pixel birth--persistence regions (part II).}
We include only distinguishing pixels for which there is (are) top node(s) within a cohort. `Left' and `Right' refer to the hemispheres of the brain. We use the following abbreviations: superior frontal gyrus medial segment (MSFG), superior temporal gyrus (STG), opercular part of the inferior frontal gyrus (OpIFG), transverse temporal gyrus (TTG), frontal operculum (FO), gyrus rectus (GRe), middle frontal gyrus (MFG), orbital part of the inferior frontal gyrus (OrIFG), precuneus (PCu), cuneus (CC), anterior insula (AIns), superior parietal lobule (SPL), lingual gyrus (LiG), cerebellum exterior (CE), parahippocampal gyrus (PHG), medial frontal cortex (MFC), medial orbital gyrus (MOrG), and posterior cingulate gyrus (PCgG). 
 }
 \label{table:features2}
 \begin{tabular}{*7c}
 \br
 Pixel Indices &  \multicolumn{2}{c}{Siblings} & \multicolumn{2}{c}{Controls} & \multicolumn{2}{c}{Patients}\\
 \mr
 {} & Node& Location & Node & Location & Node & Location\\ \mr
 14		& 	84	& Left PCu			&	112&	Left SPL		&	--	&	--	\\	 
		& 112	& Left SPL				&	--  &	--				&	--	&	--	\\
		\mr	 
15		& 51	& 	Right LiG			&	--	&	--				&	--	&	--	\\	
		\mr 
17		& --	& -- 					&112	&Left SPL			&26	& Left AIns	\\	
		\mr 		
18		&51 	& Right LiG			&	69	&Right MSFG		&	--	&--		\\	 
		\mr
19		& 	3	& Right Amyg.		&	--	&	--				& --	&	--	\\	 
		\mr
21		& 	3	& Right Amyg.		&	8	&Left CE			&	--	&	--	\\	 
		& --	& 	--					&	112&Left SPL 		&--		&--		\\	 		\mr		
22		& 	80	& Left OrIFG			&	--	&	--				&	--	&	--	\\	 
		& 	112& Left SPL			&	--	&	--				&	--	&	--	\\
		\mr	 
24		& 	80		& Left OrIFG		&	59	&Right MFG		&	--	&	--	\\	 
		& 	112	&Left SPL 			&	--	&	--				&	--	&	--	\\	 				\mr	
25		& 112	&Left SPL 				&	--	&	--				&	--	&	--	\\	
		 
 \br
 \end{tabular}
 \end{table}

   \begin{table}
 \centering
 \caption{{\bf Top nodes that are involved in loop representatives within the bounds of distinguishing pixel birth--persistence regions (part III).}
We include only distinguishing pixels for which there is (are) top node(s) within a cohort. `Left' and `Right' refer to the hemispheres of the brain. We use the following abbreviations: superior frontal gyrus medial segment (MSFG), superior temporal gyrus (STG), opercular part of the inferior frontal gyrus (OpIFG), transverse temporal gyrus (TTG), frontal operculum (FO), gyrus rectus (GRe), middle frontal gyrus (MFG), orbital part of the inferior frontal gyrus (OrIFG), precuneus (PCu), cuneus (CC), anterior insula (AIns), superior parietal lobule (SPL), lingual gyrus (LiG), cerebellum exterior (CE), parahippocampal gyrus (PHG), medial frontal cortex (MFC), medial orbital gyrus (MOrG), and posterior cingulate gyrus (PCgG). 
 }
 \label{table:features3}
 \begin{tabular}{*7c}
 \br
 Pixel Indices &  \multicolumn{2}{c}{Siblings} & \multicolumn{2}{c}{Controls} & \multicolumn{2}{c}{Patients}\\
 \midrule
 {} & Node& Location & Node & Location & Node & Location\\ \mr
 27		& 112	& Left SPL				&60	& Left MFG		&	--	&	--	\\
		\mr	 
28		& 119	& Right TTG	 		&	--	&	-- 				&	--	&	--	\\	
		\mr 
30		&119 	& Right TTG			&59	& Right MFG		&	--	&	--	\\							\mr
31		& 119	&  Right TTG			&59	& Right MFG		&--	&	--	\\
		\mr	 
32		& 120	& 	Left TTG			&	--	&	--				&	--	&	--	\\
		\mr	 
33		& --	& -- 					&	59	&	Right MFG	&	--	&	--	\\
		\mr	 
34		& 	120& Left TTG			&	8	&Left CE			&	--	&	--	\\	 	
		& --	& --					&	85	&Right PHG		&	--	&	--	\\
		\mr	
36		& --	& 	--					&	8	& Left CE 			&	--	&	--	\\	 
		& --	& 	--					&	85	& Right PHG		&	--	&	--	\\

 \bottomrule
 \end{tabular}
 \end{table}

  \begin{table}
 \centering
 \caption{{\bf Top nodes that are involved in loop representatives within the bounds of distinguishing pixel birth--persistence regions (part IV).} 
We include only distinguishing pixels for which there is (are) top node(s) within a cohort. `Left' and `Right' refer to the hemispheres of the brain. We use the following abbreviations: superior frontal gyrus medial segment (MSFG), superior temporal gyrus (STG), opercular part of the inferior frontal gyrus (OpIFG), transverse temporal gyrus (TTG), frontal operculum (FO), gyrus rectus (GRe), middle frontal gyrus (MFG), orbital part of the inferior frontal gyrus (OrIFG), precuneus (PCu), cuneus (CC), anterior insula (AIns), superior parietal lobule (SPL), lingual gyrus (LiG), cerebellum exterior (CE), parahippocampal gyrus (PHG), medial frontal cortex (MFC), medial orbital gyrus (MOrG), and posterior cingulate gyrus (PCgG). 
 }
 \label{table:features4}
 \begin{tabular}{*7c}
 \br
 Pixel Indices &  \multicolumn{2}{c}{Siblings} & \multicolumn{2}{c}{Controls} & \multicolumn{2}{c}{Patients}\\
 \mr
 {} & Node & Location & Node & Location & Node & Location\\ \mr

37		& 	57	& Right MFC			& 59	& Right MFG		&	--	&	--	\\	 				& 	58	& Left MFC			&--	&	--				&	--	&	--	\\	 
		& 	119& Right TTG			&	--	&	--				&	--	&	--	\\
		\mr	 
39		& 	111& Right SPL			&	26	&Left AIns		&	--	&	--	\\	 
		& --	& --					&	52	&Left LiG			&	--	&	--	\\	 							
		\mr
40		& 46	&Left GRe 			&	--	&	--				&	--	&	--	\\	 
		& 57	& Right MFC			&	--	&	--				&	--	&--		\\	 
		& 58	& Left MFC			&	--	&	--				&	--	&	--	\\	 
		& 64	& Left MOrG 			&	--	&	--				&	--	&	--	\\	 			 	\mr
41		& 82	& Left PCgG			&	--	&	--				&	--	&	--	\\	 
																						
 \br
 \end{tabular}
 \end{table}

 
\section{Discussion}\label{Sec:Discussion} 

We applied methods from persistent homology to analyze loops in functional brain networks of schizophrenia patients, siblings of schizophrenia patients, and healthy controls. We constructed both persistence landscapes and persistence images of these networks, and we compared them to each other using several clustering techniques.

We observed topological differences in the functional brain networks of schizophrenia patients, siblings of schizophrenia patients, and healthy controls with respect to the loops in their networks. We also found that PLs and PIs have different
practical advantages and disadvantages when applied to the same data set; these insights may be useful for interpreting the results of PH computations in networks in diverse applications.

Computing PLs gave interesting results when the comparing mean PLs of the cohorts but not when comparing individual landscapes of the subjects. Using mean PLs, we were able to separate the sibling cohort from the other two subject groups in each of the four time regimes. This is supported by the p-values that we obtained for the distances between the mean landscapes of the sibling cohort versus the controls and patient cohorts, although not all of our p-values are statistically significant. The shape of the mean PLs seems to suggest that loops that occur in the functional brain networks of siblings are more persistent on average than those in the functional networks of controls or patients. This may imply either that (1) loops in the networks of siblings tend to be longer or that (2) the third edge between three nodes has a small edge weight and thus that three brain regions with a large pairwise Pearson correlation between one region and two of the other regions do not necessarily imply that there is a large correlation between the other two brain regions; such a third edge facilitates the creation of a loop structure in the filtration. (Recall that we need at least four nodes for our loops.) To examine this issue further, it may be useful to analyze cross-links in the functional networks, as in \cite{Bassett2014II}. For the above computations and their interpretation, we need to take into account that we did not include infinitely-persisting loops (which persist until the end of a filtration). We also include only positive edge weights in our networks, so we only analyzed loops that arise from brain regions with positive pairwise Pearson correlations.

Although we were able to obtain interesting insights about the data using mean PLs, we did not find interpretable results from comparing individual landscapes, and only using the mean landscapes reduces the amount of information the we can obtain from this approach. By contrast, using individual PIs and SSVMs allowed us to separate the entire set of subjects (with $100\%$ accuracy) in each of the four time regimes. In previous work, Anderson and Cohen~\cite{Anderson2013} obtained 65\% accuracy for schizophrenia classification by applying machine-learning techniques to functional brain networks. It is important to note, however, that our results are based on using a priori knowledge of group membership (specifically, by including the maximum birth times of loops within subject groups). These birth times seem to include nontrivial information, which is important to pursue further in future studies. Moreover, such a priori knowledge is tied closely to the choice of statistical thresholding when preprocessing fMRI data. Consequently, developing a statistical model that can classify a novel subject based on a PI representation also requires further explorations of how to choose such a threshold.

Computing PIs also allowed us to identify brain regions with consistent involvement in loops in the functional networks within subject cohorts. Of the three cohorts, we found that siblings have the highest level of consistent brain-region involvement in the performance of the mental task in this study across the four time regimes. That is, brain regions that are involved in loops for siblings in one of the time regimes are more likely to also be involved in loops in other time regimes than is the case for patients or controls. It is particularly noteworthy that the number of brain regions that are consistently involved in the separation of the three cohorts is larger in the siblings of schizophrenia patients than in the healthy controls. We view heterogeneous involvement of brain regions in loops as a notion of neurological `flexibility'. Various works have studied concepts of brain flexibility using community structure~\cite{Bassett2011,Braun2016}. In those studies, flexibility was defined differently --- based on how often a brain region changes its allegiance to a community of nodes over time, so it does not use loops directly --- than in the above characterization, but it is noteworthy that Braun \emph{et al.}~\cite{Braun2016} observed that relatives of schizophrenia patients have larger flexibility than healthy controls. In our work, we found that a specific group of brain regions leads to the separation of the three subject groups when using PIs and observed for the schizophrenia patients that the regions that lead to a separation consistently in each of the four time regimes are fewer in number than for the siblings and the controls. Braun \emph{et al.}\cite{Braun2016} reported that there is larger node flexibility in network organization of schizophrenia patients than in healthy controls. Additionally, Siebenh\"uhner \emph{et al.}~\cite{Siebenhuhner2013} observed a greater variability in temporal networks that were constructed from magnetoencephalography (MEG) data of schizophrenia patients than those that were constructed from the data of healthy controls.

We observed four time regimes, which each consist of fMRI signals that were recorded during one block of a $0$-back task and a $2$-back task. For time regime 2, we obtained very small and statistically significant p-values in our mean PL computations when comparing siblings to controls and when we were comparing siblings to a group that consists of all patients and all controls. We did not observe this for any of the other time regimes. We conclude that time regime 2 appears to capture significant changes in the persistence and/or appearance of loops in the networks of siblings during the working-memory task. It will be useful to conduct further {laboratory} experiments to draw biological conclusions.

Schizophrenia has a high genetic determinism, so siblings of schizophrenia patients have a significant genetic risk of developing the disease themselves~\cite{Bertolino2009}, and it has been demonstrated that they have abnormalities in their structural neuronal networks~\cite{Collin2014}. 
Although our results that functional brain networks that are constructed from fMRI measurements of siblings differ both from those of patients and those of healthy controls do not agree completely with the current scientific literature, other studies have also reported that the features of fMRIs of siblings of schizophrenia patients differ both from schizophrenia patients and from controls. For example, Callicott \emph{et al.}~\cite{Callicott2003} observed in an fMRI study that there was no difference in task performance between healthy siblings of schizophrenia patients and healthy controls, yet they detected a physiological similarity between the sibling cohort and the schizophrenia patients in the associated fMRI data. Similarly, Sepede \emph{et al.}~\cite{Sepede2010} observed using fMRI data from a different data set that healthy siblings of schizophrenia patients exhibit differences in brain function from schizophrenia patients, although they did not differ significantly in task performance. Additionally, it was demonstrated recently that schizophrenia patients undergo a cortical normalization process over the course of the disease~\cite{Guo2016}. However, one needs further phenotypic information to assess whether it is possible to directly connect the results of such a study to our observations.

Because our results are somewhat inconsistent with prior observations, it is also possible that our data set contains experimental noise that is beyond our control. Using standard network-analysis techniques on the functional brain networks, we did not observe any differences between the three subject groups. Nevertheless, we believe that our comparison of PLs to PIs and the different types of results from these techniques provide a valuable example of an approach that uses topological data analysis for functional brain networks. To give another cautionary note, one needs to take into account that there are difficulties when interpreting the information about node participation in loops from computations of PH, as the software that is used for such computations (including, specifically, {\sc javaPlex}, which is what we used) only finds representatives of the loops. These representatives are not determined in an optimal way, and they need not be `geometrically nice'~\cite{Adams2014}. For example, in these calculations, one often encounters double loops or even triple loops as generators for one loop in a functional network. Selecting a basis of homology generators that behaves in a biologically representative way corresponds mathematically to solving a problem known as the `optimal homology-basis problem', which is difficult (and is NP-hard in the worst case)~\cite{Erickson2000}. Despite these difficulties, our list of discriminating nodes provides a useful starting point for further investigations into neuronal abnormalities in functional networks of schizophrenia patients.

Another important issue is that we preprocessed the data for our study. This is very common when working with fMRI data, but such steps are not uncontroversial, and studies on functional connectivity in schizophrenia patients have found contradictory results depending on whether one performs global signal correction~\cite{Fox2009,Fornito2015}. It is also relevant to keep in mind that the choice of functional connectivity measure can influence results~\cite{Smith2011}. We used the Pearson correlation because of its simplicity and the fact that it is a widely used measure of functional connectivity~\cite{Wang2010,Bassett2012}. Many other choices are also available.

Finally, we chose to threshold our networks and removed edges whose weights were below a certain amount. However, it has been observed previously \cite{Bassett2012} that edges with small weights can be important when comparing functional brain networks of schizophrenia patients to those of healthy controls. However, such missing edges --- depending on their location and distribution --- can result in loops in a network, our analysis indirectly includes some of this information. In future work, it seems interesting to consider only the parts of the functions networks that were below the threshold value that we employed in our present study and analyze them with PH to compare classification results.
   

\section*{Acknowledgements}
\label{sec:Acknowledgements}

We thank Alessandro Bertolino, Fabio Sambataro, and the Bari psychiatric neuroscience group for permission to study their data. All rights of the data lie with their research groups, and we are unable to release the data. We also thank Pawel D\l otko for his help with the {\sc Persistence Landscapes Toolbox} and for providing us with new versions of the code during our project. We are also grateful to Danielle Bassett, Peter Bubenik, Carina Curto, Parker Edwards, and several referees for helpful comments. We thank Florian Lipsmeier and Franziska Mech from Roche for useful discussions. We also acknowledge Advanced Research Computing (ARC) at University of Oxford for resources that we used in carrying out this work. BJS thanks the EPSRC and MRC (grant number EP/G037280/1) and F. Hoffmann-La Roche AG for funding her doctoral studies. HAH acknowledges funding from an EPSRC Fellowship (EP/K041096/1) and a Royal Society University Research Fellowship. BJS and HAH are members of the Centre for Topological Data Analysis, which is funded by an EPSRC grant (EP/R018472/1).



\begin{thebibliography}{100}
\expandafter\ifx\csname url\endcsname\relax
  \def\url#1{{\tt #1}}\fi
\expandafter\ifx\csname urlprefix\endcsname\relax\def\urlprefix{URL }\fi
\providecommand{\eprint}[2][]{\url{#2}}

\bibitem{WHOSchizophrenia}
{World Health Organization} 19 September 2015 Schizophrenia
  \url{http://www.who.int/mental\_health/management/schizophrenia/en/}

\bibitem{Bertolino2009II}
Bertolino A and Blasi G 2009 {\em Neuroscience\/} {\bf 164} 288--299

\bibitem{Dawson2014}
Dawson N, Xiao X, McDonald M, Higham D~J, Morris B~J and Pratt J~A 2014 {\em
  Cerebral Cortex\/} {\bf 24} 452--464

\bibitem{Bullmore1997}
Bullmore E~T, Frangou S and Murray R~M 1997 {\em Schizophrenia Research\/} {\bf
  28} 143--156

\bibitem{Peled1999}
Peled A, Geva A~B, Kremen W~S, Blankfeld H~M, Esfandiarfard R and Nordahi T~E
  2001 {\em International Journal of Neuroscience\/} {\bf 106} 47--61

\bibitem{Bassett2008}
Bassett D~S, Bullmore E~T, Verchinski B~A, Mattay V~S, Weinberger D~R and
  Meyer-Lindenberg A 2008 {\em The Journal of Neuroscience\/} {\bf 28}
  9239--9248

\bibitem{Fornito2012}
Fornito A, Zalesky A, Pantelis C and Bullmore E~T 2012 {\em NeuroImage\/} {\bf
  62} 2296--2314

\bibitem{Zalesky2012}
Zalesky A, Fornito A, Egan G~F, Pantelis C and Bullmore E~T 2012 {\em Human
  Brain Mapping\/} {\bf 33} 2535--2549

\bibitem{Fornito2015}
Fornito A and Bullmore E~T 2015 {\em Current Opinion in Neurobiology\/} {\bf
  30} 44--50

\bibitem{Bullmore2009}
Bullmore E~T and Sporns O 2009 {\em Nature Reviews\/} {\bf 10} 186--198

\bibitem{Bullmore2011}
Bullmore E~T and Bassett D~S 2011 {\em Annual Review of Clinical Psychology\/}
  {\bf 7} 113--140

\bibitem{Sporns2014}
Sporns O 2014 {\em Nature Reviews Neuroscience\/} {\bf 17} 652--660

\bibitem{Papo2014}
Papo D, Zanin M, Pineda-Pardo J~A, Boccaletti S and Buld\'{u} J~M 2014 {\em
  Philosophical Transactions of the Royal Society B\/} {\bf 369} 20130525

\bibitem{Papo2014II}
Papo D, Buld\'{u} J~M, Boccaletti S and Bullmore E~T 2014 {\em Philosophical
  Transactions of the Royal Society B\/} {\bf 369} 20130520

\bibitem{Betzel2016}
Betzel R~F and Bassett D~S 2017 {\em NeuroImage\/} {\bf 160} 73--83 ISSN
  1053-8119 functional Architecture of the Brain

\bibitem{bassett2017}
Bassett D~S and Sporns O 2017 {\em Nature Neuroscience\/} {\bf 20} 353--364

\bibitem{bassett2018}
Bassett D~S, Zurn P and Gold J~I 2018 {\em Nature Reviews Neuroscience\/} {\bf
  19} 566--578

\bibitem{Sporns2015}
Sporns O 2015 Graph-theoretical analysis of brain networks {\em Brain Mapping:
  An Encyclopedic Reference\/} vol~1 ed Toga A~W (Cambridge, Massachusetts:
  Academic Press: Elsevier) pp 629--633

\bibitem{Petersen2015}
Petersen S~E and Sporns O 2015 {\em Neuron\/} {\bf 88} 207--219

\bibitem{Stolz2017}
Stolz B~J, Harrington H~A and Porter M~A 2017 {\em Chaos\/} {\bf 27} 047410

\bibitem{eklund2016}
Eklund A, Nichols T~E and Knutsson H 2016 {\em Proceedings of the National
  Academy of Sciences of the United States of America\/} {\bf 113} 7900--7905

\bibitem{Lynall2010}
Lynall M~E, Bassett D~S, Kerwin R, McKenna P~J and Kitzbichler M 2010 {\em The
  Journal of Neuroscience\/} {\bf 30} 9477--9487

\bibitem{Rubinov2013}
Rubinov M and Bullmore E~T 2013 {\em Dialogues in Clinical Research\/} {\bf 15}
  339--349

\bibitem{Alexander-Bloch2012}
Alexander-Bloch A~F, Lambiotte R, Roberts B, Giedd J, Gogtay N and Bullmore E~T
  2012 {\em NeuroImage\/} {\bf 15} 3889--3900

\bibitem{Liu2008}
Liu Y, Linag M, Zhou Y, He Y, Hao Y, Song M, Yu C, Liu H, Liu Z and Jiang T
  2008 {\em Brain\/} {\bf 131} 945--961

\bibitem{Singh2016}
Singh M and Bagler G 2016 {\em arXiv:1602.01191\/}

\bibitem{flanagan2018}
Flanagan R, Lacasa L, Towlson E~K, Lee S~H and Porter M~A 2019 {\em Journal of
  Complex Networks\/} {\bf 7} 932--960

\bibitem{Alexander-Bloch2010}
Alexander-Bloch A~F, Gogtay N, Meunier D, Birn R, Clasen L, Lalonde F, Lenroot
  R, Giedd J and Bullmore E~T 2010 {\em Frontiers in Systems Neuroscience\/}
  {\bf 4} 1--16

\bibitem{towlson2019}
Towlson E~K, V\'{e}rtes P~E, M\"{u}ller U and Ahnert S~E 2019 {\em Frontiers in
  Psychiatry\/} {\bf 10} 611

\bibitem{Edelsbrunner2002}
Edelsbrunner H, Letscher D and Zomorodian A 2002 {\em Discrete and
  Computational Geometry\/} {\bf 28} 511--533

\bibitem{Edelsbrunner2008}
Edelsbrunner H and Harer J~L 2008 Persistent homology --- {A} survey {\em
  Surveys on Discrete and Computational Geometry. Twenty years later\/} ({\em
  Contemporary Mathematics\/} vol 453) ed Goodman J~E, Pach J and Pollak R
  (American Mathematical Society) pp 257--282

\bibitem{Ghrist2008}
Ghrist R 2008 {\em Bulletin of the American Mathematical Society\/} {\bf 45}
  61--75

\bibitem{Carlsson2009}
Carlsson G 2009 {\em Bulletin of the American Mathematical Society\/} {\bf 46}
  255--308

\bibitem{Edelsbrunner2010}
Edelsbrunner H and Harer J~L 2010 {\em Computational Topology\/} (Providence R.
  I.: American Mathematical Society)

\bibitem{sizemore2018}
Sizemore A~E, Phillips-Cremins J~E, Ghrist R and Bassett D~S 2018 {\em Network
  Neuroscience\/} {\bf 3} 656--673

\bibitem{batt2020}
Battiston F, Cencetti G, Iacopini I, Latora V, Lucas M, Patania A, Young J~G
  and Petri G 2020 {\em Physics Reports\/} ISSN 0370-1573
  \urlprefix\url{http://www.sciencedirect.com/science/article/pii/S0370157320302489}

\bibitem{Newman2018}
Newman M~E~J 2018 {\em Networks\/} 2nd ed (Oxford, UK: Oxford University Press)

\bibitem{Bollobas1998}
Bollob\'{a}s B 1998 {\em Modern Graph Theory\/} (Heidelberg, Germany:
  Springer-Verlag)

\bibitem{Bassett2014}
Bassett D~S, Yang M, Wymbs N~F and Grafton S~T 2015 {\em Nature Neuroscience\/}
  {\bf 18} 744--751

\bibitem{otter2017}
Otter N, Porter M~A, Tillmann U, Grindrod P and Harrington H~A 2017 {\em
  European Physical Journal --- Data Science\/} {\bf 6} 1--38

\bibitem{Patania2017}
Patania A, Vaccarino F and Petri G 2017 {\em European Physical Journal --- Data
  Science\/} {\bf 6} 7

\bibitem{kramar2013}
Kram\'ar M, Goullet A, Kondic L and Mischaikow K 2013 {\em Physical Review E\/}
  {\bf 87} 042207

\bibitem{taylor2015}
Taylor D, Klimm F, Harrington H~A, Kram\'ar M, Mishchaikow K, Porter M~A and
  Mucha P~J 2015 {\em Nature Communications\/} {\bf 6} 7723

\bibitem{bgk2015}
Bhattacharya S, Ghrist R and Kumar V 2015 {\em IEEE Transactions on Robotics\/}
  {\bf 31} 578--590

\bibitem{topaz2014}
Topaz C~M, Ziegelmeier L and Halverson T 2015 {\em PLoS ONE\/} {\bf 10}
  e0126383

\bibitem{Bendich2014}
Bendich P, Marron J~S, Miller E, Pieloch A and Skwerer S 2016 {\em Annals of
  Applied Statistics\/} {\bf 10} 198--218

\bibitem{Feng2020}
Feng M and Porter M~A Spatial applications of topological data analysis:
  {C}ities, snowflakes, random structures, and spiders spinning under the
  influence arXiv:2001.01872 (Physical Review Research, in press)

\bibitem{Byrne2019}
Byrne H~M, Harrington H~A, Muschel R, Reinert G, Stolz B~J and Tillmann U 2019
  {\em Mathematics Today\/} {\bf 55} 206--210

\bibitem{Curto2008}
Curto C and Itskov V 2008 {\em PLoS Computational Biology\/} {\bf 4} e000205

\bibitem{Dabaghian2012}
Dabaghian Y, M\'{e}moli F, Frank L and Carlsson G~E 2012 {\em PLoS
  Computational Biology\/} {\bf 8} e1002581

\bibitem{Petri2013}
Petri G, Scolamiero M, Donato I and Vaccarino F 2013 {\em PLoS ONE\/} {\bf 8}
  e66505

\bibitem{Lee2011}
Lee H, Chung M~K, Kang H, Kim B~N and Lee D~S 2011 Discriminative persistent
  homology of brain networks {\em IEEE International Symposium on Biomedical
  Imaging: {F}rom Nano to Macro\/} pp 841--844

\bibitem{Giusti2015}
Giusti C, Pastalkova E, Curto C and Itskov V 2015 {\em Proceedings of the
  National Academy of Sciences of the United States of America\/} {\bf 112}
  13455--13460

\bibitem{Spreemann2015}
Spreemann G, Dunn B, Botnan M~B and Baas N~A 2018 {\em Physical Review E\/}
  {\bf 97} 032313

\bibitem{Curto2016}
Curto C 2017 {\em Bulletin of the American Mathematical Society\/} {\bf 54}
  63--78

\bibitem{Giusti2016}
Giusti C, Ghrist R and Bassett D~S 2016 {\em Journal of Computational
  Neuroscience\/} {\bf 41} 1--14

\bibitem{Reimann2017}
Reimann M~W, Nolte M, Scolamiero M, Turner K, Perin R, Chindemi G, D{\l}otko P,
  Levi R, Hess K and Markram H 2017 {\em Frontiers in Computational
  Neuroscience\/} {\bf 11} 48

\bibitem{Dabaghian2016}
Babichev A and Dabaghian Y 2017 Persistent memories in transient networks {\em
  Emergent Complexity from Nonlinearity, in Physics, Engineering and the Life
  Sciences\/} (Springer-Verlag) pp 179--188

\bibitem{Lee2019}
Lee H, Chung M~K, Choi H, Kang H, Ha S, Kim Y~K and Lee D~S 2019 Harmonic holes
  as the submodules of brain network and network dissimilarity {\em
  Computational Topology in Image Context. CTIC 2019\/} ({\em Lecture Notes in
  Computer Science\/} vol 11382) ed Marfil R, Calder{\'o}n M, del R{\'\i}o F~D
  and P~Real A~B (Cham, Switzerland: Springer International Publishing) pp
  110--122

\bibitem{Bardin2018}
Bardin J~B, Spreemann G and Hess K 2018 {\em Network Neuroscience\/} {\bf 3}
  725--743

\bibitem{chung2019}
Chung M~K, Lee H, Ombao H and Solo V 2019 {\em Network Neuroscience\/} {\bf 3}
  674--694

\bibitem{babichev2018}
Babichev A, Morozov D and Dabaghian Y 2018 {\em Network Neuroscience\/} {\bf 3}
  707--724

\bibitem{geniesse2019}
Geniesse C, Sporns O, Petri G and Saggar M 2019 {\em Network Neuroscience\/}
  {\bf 3} 763--778

\bibitem{Ibanez2019}
Ib{\'a}{\~n}ez-Marcelo E, Campioni L, Phinyomark A, Petri G and Santarcangelo
  E~L 2019 {\em NeuroImage\/} {\bf 200} 437--449

\bibitem{Croom}
Croom F~H 1978 {\em Basic Concepts of Algebraic Topology\/} (Heidelberg,
  Germany: Springer-Verlag)

\bibitem{Talairach1988}
Talairach J and Tournoux P 1988 {\em Co-Planar Stereotaxic Atlas of the Human
  Brain. {3-D} Proportional System: {A}n Approach to Cerebral Imaging\/} (New
  York City, NY, USA: Thieme Medical Publishers)

\bibitem{Bertolino2010}
Bertolino A, Taurisano P, Pisciotta N~M, Blasi G, Fazio L, Romano R, Gelao B,
  Biancho L~L, Lozupone M, Giorgio A~D, Grazia, Sambataro F, Niccoli-Asabella
  A, Papp A, Ursini G, Sinibaldi L, Popoloizo T, Sadee W and Rubini G 2010 {\em
  PLoS one\/} {\bf 5} e9348

\bibitem{Sambataro2009}
Sambataro F, Blasi G, Fazio L, Caforio G, Taurisano P, Romano R, Giorgio A~D,
  Gelao B, Biancho L~L, Papazacharias A, Popolizio T, Nardini M and Bertolino A
  2010 {\em Neuropsychopharmacology\/} {\bf 35} 904--912

\bibitem{Rampino2014}
Rampino A, Walker R~M, Torrance H~S, Anderson S~M, Fazio L, Di~Giorgio A,
  Taurisano P, Gelao B, Romano R, Masellis R, Ursini G, Caforio G, Blasi G,
  Millar J~K, Porteous D~J, Thomson P~A, Bertolino A and Evans K~L 2014 {\em
  PLoS ONE\/} {\bf 9} e99892

\bibitem{Weissenbacher2009}
Weissenbacher A, Kasess C, Gerstl F, Lanzenberger R, Moser E and Windischberger
  C 2009 {\em NeuroImage\/} {\bf 47} 1408--1416

\bibitem{Dagli1999}
Dagli M~S, Ingeholm J~E and Haxby J~V 1999 {\em NeuroImage\/} {\bf 9} 407--415

\bibitem{Birn2006}
Birn R~M, Diamond J~B, Smith M~A and Bandettini P~A 2006 {\em NeuroImage\/}
  {\bf 31} 1536--1548

\bibitem{Fox2007}
Fox M~D and Raichle M~E 2007 {\em Nature Reviews Neuroscience\/} {\bf 8} 700

\bibitem{Friston1996}
Friston K~J, Williams S, Howard R, Frackowiak R~S~J and Turner R 1996 {\em
  Magnetic Resonance in Medicine\/} {\bf 35} 346--355

\bibitem{Murphy2009}
Murphy K, Birn R~M, Handwerker D~A, Jones T~B and Bandettini P~A 2009 {\em
  NeuroImage\/} {\bf 44} 893--905

\bibitem{Fox2009}
Fox M~D, Zhang D, Snyder A~Z and Raichle M~E 2009 {\em Journal of
  Neurophysiology\/} {\bf 101} 3270--3283

\bibitem{Smith2011}
Smith S~M, Miller K~L, Salimi-Khorshidi G, Webster M, Beckmann C~F, Nichols
  T~E, Ramsay J~D and Woolrich M~W 2011 {\em NeuroImage\/} {\bf 54} 875--891

\bibitem{Zhou2009}
Zhou D, Thompson W~K and Siegle G 2009 {\em NeuroImage\/} {\bf 47} 1590--1607

\bibitem{Bassett2011}
Bassett D~S, Wymbs N~F, Porter M~A, Mucha P~J, Carlson J~M and Grafton S~T 2011
  {\em Proceedings of the National Academy of Sciences of the United States of
  America\/} {\bf 108} 7641--7646

\bibitem{Kosniowski1980}
Kosniowski C 1980 {\em A First Course in Algebraic Topology\/} (Cambridge, UK:
  Cambridge University Press)

\bibitem{feng2019}
Feng M and Porter M~A Persistent homology of geospatial data: {A} case study
  with voting arXiv:1902.05911 (SIAM Review, in press)

\bibitem{Lee2012}
Lee H, Kang H, Chung M~K, Kim B~N and Lee D~S 2012 Weighted functional brain
  network modeling via network filtration {\em NIPS Workshop on Algebraic
  Topology and Machine Learning\/}

\bibitem{Petri2014}
Petri G, Expert P, Turkheimer F, Carhart-Harris R, Nutt D, Hellyer P~J and
  Vaccarino F 2014 {\em Journal of the Royal Society Interface\/} {\bf 11}
  20140873

\bibitem{Bubenik2015I}
Bubenik P 2015 {\em Journal of Machine Learning Research\/} {\bf 16} 77--102

\bibitem{Bubenik2015}
Bubenik P and D{\l}otko P 2017 {\em Journal of Symbolic Computation\/} {\bf 78}
  91--114

\bibitem{Adams2015}
Adams H, Chepushtanova S, Emerson T, Hanson E, Kirby M, Motta F, Neville R,
  Peterson C, Shipman P and Ziegelmeier L 2017 {\em Journal of Machine Learning
  Research\/} {\bf 18} 218--252

\bibitem{Carlsson2005}
Carlsson G, Zomorodian A, Collins A and Guibas L~J 2005 {\em International
  Journal of Shape Modeling\/} {\bf 11} 149--187

\bibitem{Cohen-Steiner2005}
Cohen-Steiner D, Edelsbrunner H and Harer J 2007 {\em Discrete Computational
  Geometry\/} {\bf 27} 103--120

\bibitem{Kovacev2015}
Kovacev-Nikolic V, Bubenik P, Nikolic D and Heo G 2016 {\em Statistical
  Applications in Genetics and Molecular Biology\/} {\bf 15} 1--27

\bibitem{Wang2015}
Wang Y, Ombao H and Chung M~K 2015 Topological seizure origin detection in
  electroencephalographic signals {\em IEEE 12th International Symposium on
  Biomedical Imaging (ISBI)\/} pp 351--354

\bibitem{Dlotko2016II}
D{\l}otko P and Wanner T 2016 {\em Physica D\/} {\bf 334} 60--81

\bibitem{Garg2017}
Garg A, Lu D, Popuri K and Beg M~F 2017 Brain geometry persistent homology
  marker for parkinson's disease {\em IEEE 14th International Symposium on
  Biomedical Imaging (ISBI 2017)\/} (IEEE) pp 525--528

\bibitem{Liu2016}
Liu J~Y, Jeng S~K and Yang Y~H 2016 Applying topological persistence in
  convolutional neural network for music audio signals arXiv:1608.07373

\bibitem{Kanari2019}
Kanari L, Ramaswamy S, Shi Y, Morand S, Meystre J, Perin R, Abdellah M, Wang Y,
  Hess K and Markram H 2019 {\em Cerebral Cortex\/} {\bf 29} 1719--1735

\bibitem{Kanari2018}
Kanari L, D{\l}otko P, Scolamiero M, Levi R, Shillcock J, Hess K and Markram H
  2018 {\em Neuroinformatics\/} {\bf 16} 3--13

\bibitem{PIcode}
Adams H, Chepushtanova S, Emerson T, Hanson E, Kirby M, Motta F, Neville R,
  Peterson C, Shipman P and Ziegelmeier L 2016 Persistence images
  \url{https://github.com/CSU-TDA/PersistenceImages}

\bibitem{javaPlex}
Adams H, Tausz A and Vejdemo-Johansson M 2014 {\sc JavaPlex}: {A} research
  software package for persistent (co)homology (2011) {\em Mathematical
  Software---ICMS 2014\/} vol 8592 ed Hong H and Yap C pp 129--136 software
  available at \url{http://javaplex.github.io/}

\bibitem{bron1973}
Bron C and Kerbosch J 1973 {\em Communications of the ACM\/} {\bf 16} 575--577

\bibitem{Wildmann2011}
Wildmann J 2011 Bron--{K}erbosch maximal clique finding algorithm code
  available at:
  \url{http://www.mathworks.co.uk/matlabcentral/fileexchange/30413-bron-kerbosch-maximal-clique-finding-algorithm}

\bibitem{siamcluster}
Gan G, Ma C and Wu J 2007 {\em Data Clustering: Theory, Algorithms, and
  Applications\/} (Philadelphia, PA, USA: Society for Industrial and Applied
  Mathematics)

\bibitem{Porter2009}
Porter M~A, Onnela J~P and Mucha P~J 2009 {\em Notices of the American
  Mathematical Society\/} {\bf 56} 1082--1097, 1164--1166

\bibitem{fortunato2016}
Fortunato S and Hric D 2016 {\em Physics Reports\/} {\bf 659} 1--44

\bibitem{MuchaComm}
Jeub L~G~S, Bazzi M, Jutla I~S and Mucha P~J 2011--2016 A generalized {L}ouvain
  method for community detection implemented in {M}{\sc atlab}, version 2.0
  \url{https://github.com/GenLouvain/GenLouvain}

\bibitem{Mucha2010}
Mucha P~J, Richardson T, Macon K, Porter M~A and Onnela J~P 2010 {\em
  Science\/} {\bf 328} 876--878

\bibitem{blondel2008}
Blondel V~D, Guillaume J~L, Lambiotte R and Lefebvre E 2008 {\em J. Stat. Mech:
  Theory Exp.\/} {\bf 2008} P10008

\bibitem{Bradley1998}
Bradley P~S and Mangasarian O~L 1998 Feature selection via concave minimization
  and support vector machines {\em Machine Learning Proceedings of the
  Fifteenth International Conference\/} ICML 1998 pp 82--90

\bibitem{Zhu2003}
Zhu J, Rosset S, Hastie T and Tibshirani R 2004 {\em Advances in Neural
  Information Processing Systems\/} {\bf 16} 49--56

\bibitem{Zhang2010b}
Zhang L and Zhou W 2010 {\em Neural Networks\/} {\bf 23} 373--385

\bibitem{chepushtanova2014}
Chepushtanova S, Gittins C and Kirby M 2014 Band selection in hyperspectral
  imagery using sparse support vector machines {\em Algorithms and Technologies
  for Multispectral, Hyperspectral, and Ultraspectral Imagery XX\/} vol 9088
  (International Society for Optics and Photonics) p 90881F

\bibitem{patrangenaru}
Patrangenaru V, Bubenik P, Paige R~L and Osborne D 2018 {\em arXiv preprint
  arXiv:1804.10255\/}

\bibitem{Sepede2010}
Sepede G, Ferretti A, Perrucci M~G, Gambi F, Di~Donato F, Nuccetelli F,
  Del~Gratta C, Tartaro A, Salerno R~M, Ferro F~M and Romani G~L 2010 {\em
  NeuroImage\/} {\bf 49} 1080--1090

\bibitem{Bassett2014II}
Bassett D~S, Wymbs N~F, Porter M~A, Mucha P~J and Grafton S~T 2014 {\em
  Chaos\/} {\bf 24} 013112

\bibitem{Anderson2013}
Anderson A and Cohen M~S 2013 {\em Frontiers in Human Neuroscience\/} {\bf 7}
  1--18

\bibitem{Braun2016}
Braun U, Sch{\"a}fer A, Bassett D~S, Rausch F, Schweiger J, Bilek E, Erk S,
  Romanczuk-Seiferth N, Grimm O, Haddad L, Otto K, Mohnke S, Heinz A, Zink M,
  Walter H, Meyer-Lindenberg A and Tost H 2016 {\em Proceedings of the National
  Academy of Sciences of the United States of America\/} {\bf 113} 12568--12573

\bibitem{Siebenhuhner2013}
Siebenh{\"u}hner F, Weiss S~A, Coppola R, Weinberger D~R and Bassett D~S 2013
  {\em PLoS ONE\/} {\bf 8} e72351

\bibitem{Bertolino2009}
Bertolino A, Fazio L, Giorgio A~D, Blasi G, Romano R, Taurisano P, Caforio G,
  Sinibaldi L, Ursini G, Popoloizo T, Tirotta E, Papp A, Dallapiccola B,
  Borrelli E and Sadee W 2009 {\em The Journal of Neuroscience\/} {\bf 29}
  1224--1234

\bibitem{Collin2014}
Collin G, Kahn R~S, de~Reus M~A, Cahn W and van~den Heuvel M~P 2014 {\em
  Schizophrenia Bulletin\/} {\bf 40} 438--448

\bibitem{Callicott2003}
Callicott J~H, Egan M~F, Mattay V~S, Bertolino A, Bone A~D, Verchinksi B and
  Weinberger D~R 2003 {\em American Journal of Psychiatry\/} {\bf 160} 709--719

\bibitem{Guo2016}
Guo S, Palaniyappan L, Liddle P~F and Feng J 2016 {\em Psychological
  Medicine\/} {\bf 46} 2201--2214

\bibitem{Adams2014}
Adams H and Tausz A 2015 {\sc javaPlex} tutorial ({PDF} version available
  online:
  \url{http://javaplex.googlecode.com/svn/trunk/reports/javaplex_tutorial/javaplex_tutorial.pdf})

\bibitem{Erickson2000}
Erickson J 2012 Combinatorial optimization of cycles and bases {\em Advances in
  Applied and Computational Topology\/} vol~70 ed Zomorodian A (American
  Mathematical Society, Providence, RI, USA) pp 195--228

\bibitem{Wang2010}
Wang L, Metzak P~D, Honer W~G and Woodward T~S 2010 {\em The Journal of
  Neuroscience\/} {\bf 30} 13171--13179

\bibitem{Bassett2012}
Bassett D~S, Nelson B~G, Mueller B~A, Camchong J and Lim K~O 2012 {\em
  NeuroImage\/} {\bf 59} 2196--2207

\end{thebibliography}


\providecommand{\newblock}{}


\newpage

\appendix

\section{Appendix}\label{S1}

We now give some additional details about a few of our calculations and results.


\subsection{Betti curves} \label{S1_3}

\begin{figure}[h!]
\centering
\subcaptionbox{}{\includegraphics[height=0.4\textwidth]{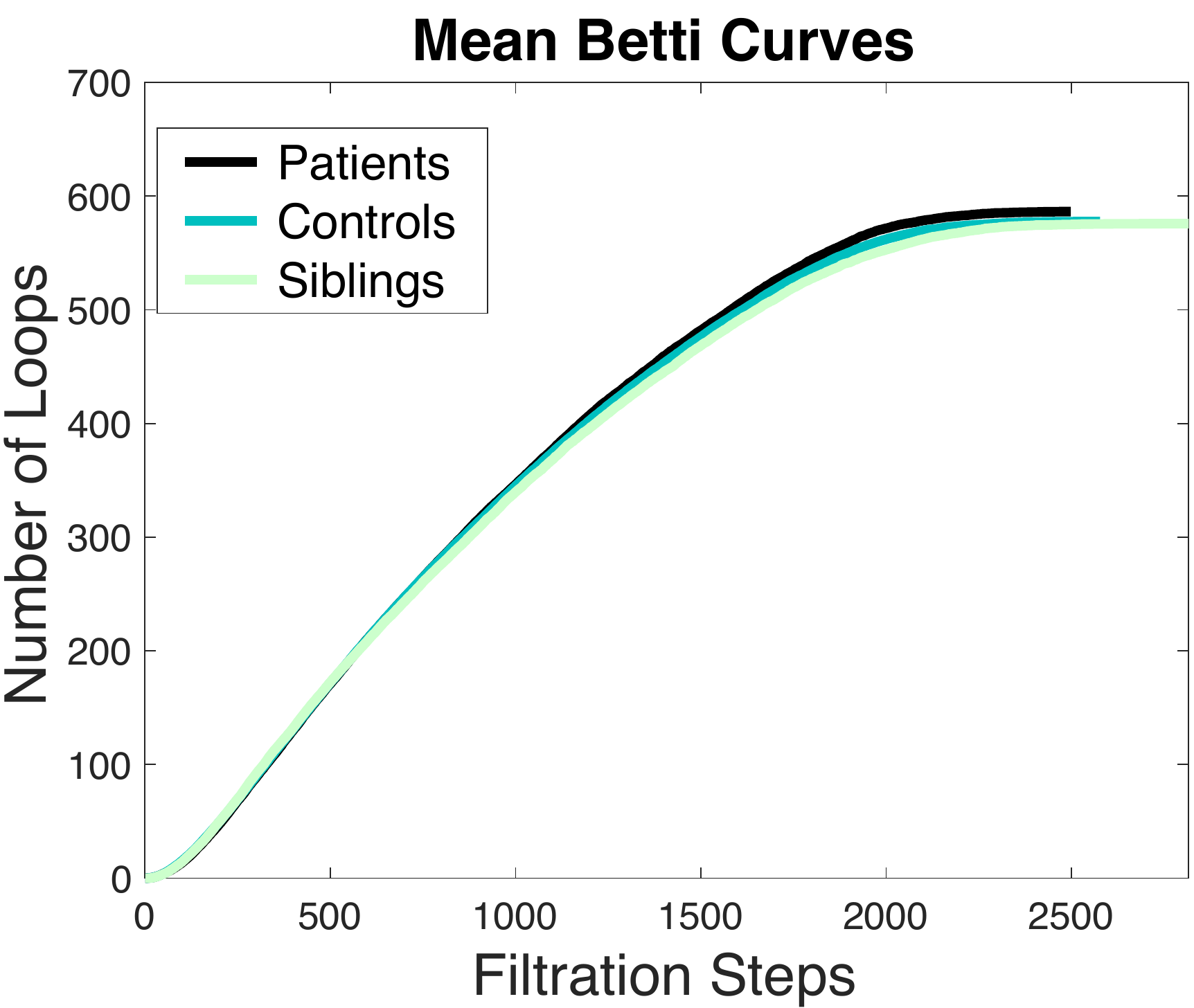}}%
\hspace{0.01\textwidth}
\subcaptionbox{}{\includegraphics[height=0.4\textwidth]{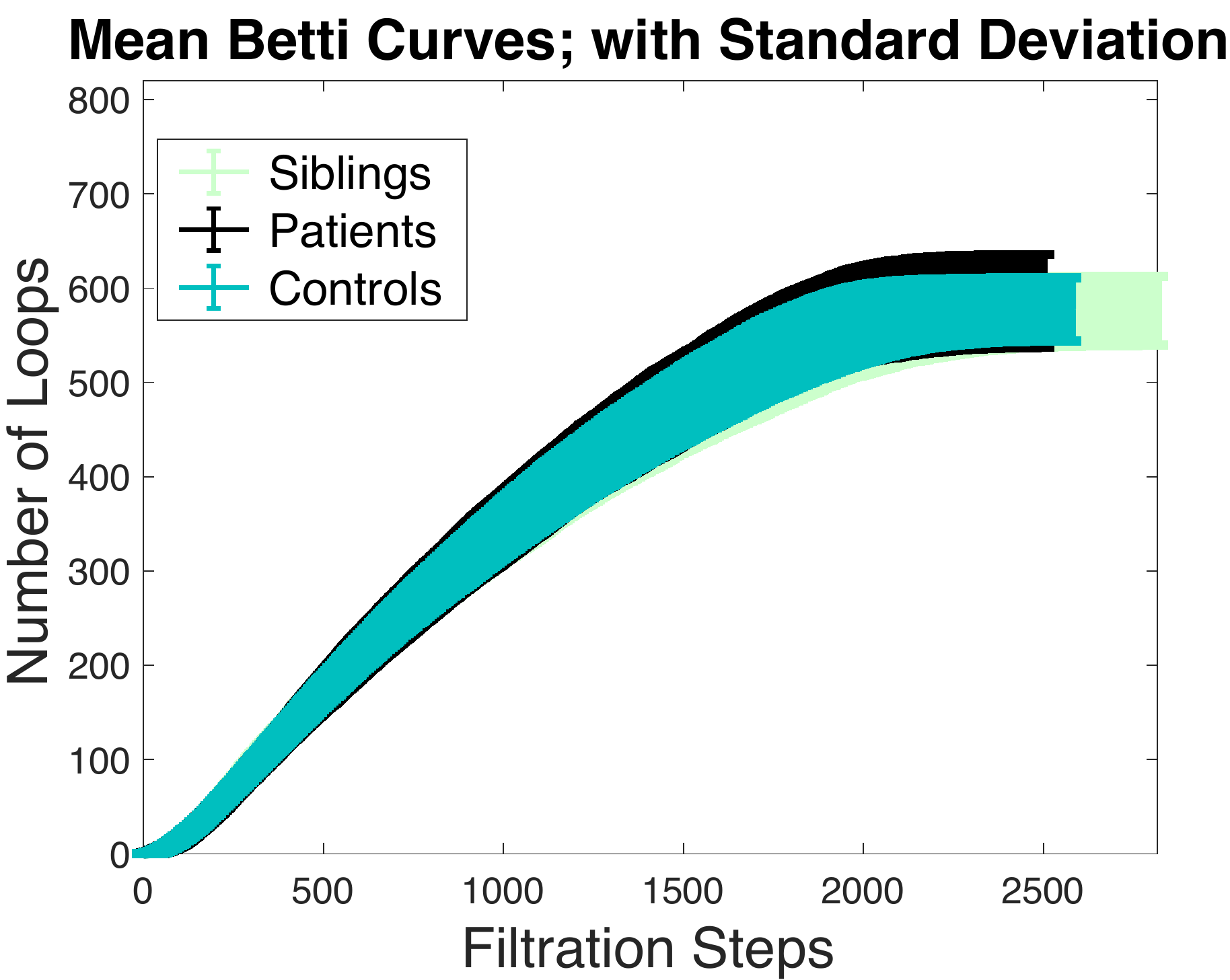}}%
\caption{(a) Mean Betti curves for the patients, controls, and siblings. (b) Mean Betti curves and their standard deviations for patients, controls, and siblings.
}\label{fig:BettiCurves}
\end{figure}

We also study \emph{Betti curves}, which were introduced in~\cite{Giusti2015} and describe how Betti numbers change across a filtration. We use each subject's full time series, which consists of 120 time points, to construct a single functional network for each subject (i.e., one time regime, rather than four separate ones). 
In all other respects, we construct the functional networks as we described in Section~\ref{construct}. We compute the mean and standard deviation across the Betti numbers for dimension $1$ (i.e., the number of loops) for each cohort in each filtration step. Aside from a slightly larger standard deviation in the patient cohort, we find that the Betti curves of the three groups look essentially the same. We show our results of computing Betti curves in Fig.~\ref{fig:BettiCurves}.

\subsection{Top brain regions in the distinguishing pixel birth--persistence bounds} \label{S1_1}

\begin{figure}[h!]
\begin{centering}
\includegraphics[width=\textwidth]{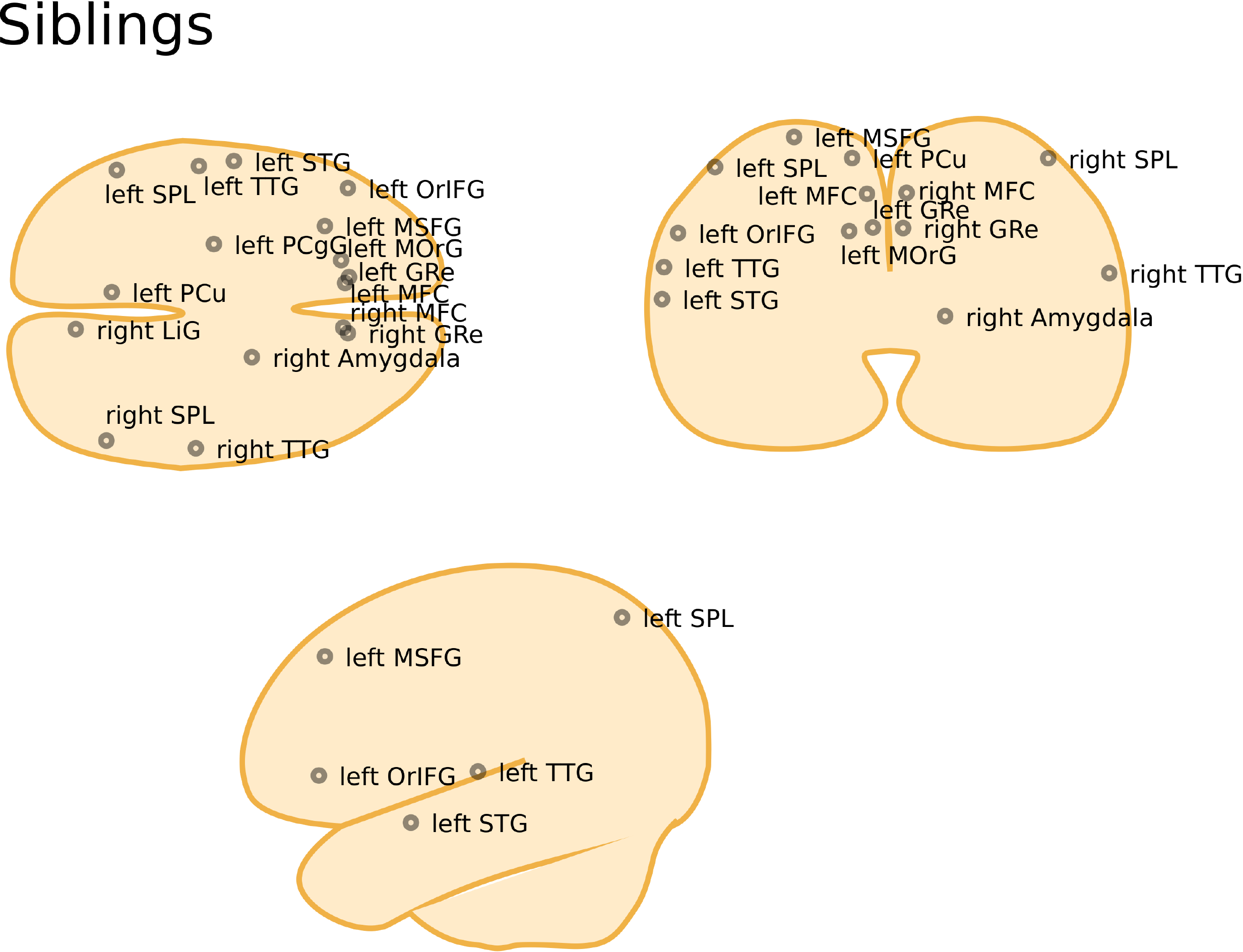}
\caption{The top nodes in representatives of loops in the distinguishing pixel birth--persistence bounds for siblings.}
\label{fig:MAX_nodes_Sib}
\end{centering}
\end{figure}

\begin{figure}[h!]
\begin{centering}
\includegraphics[width=\textwidth]{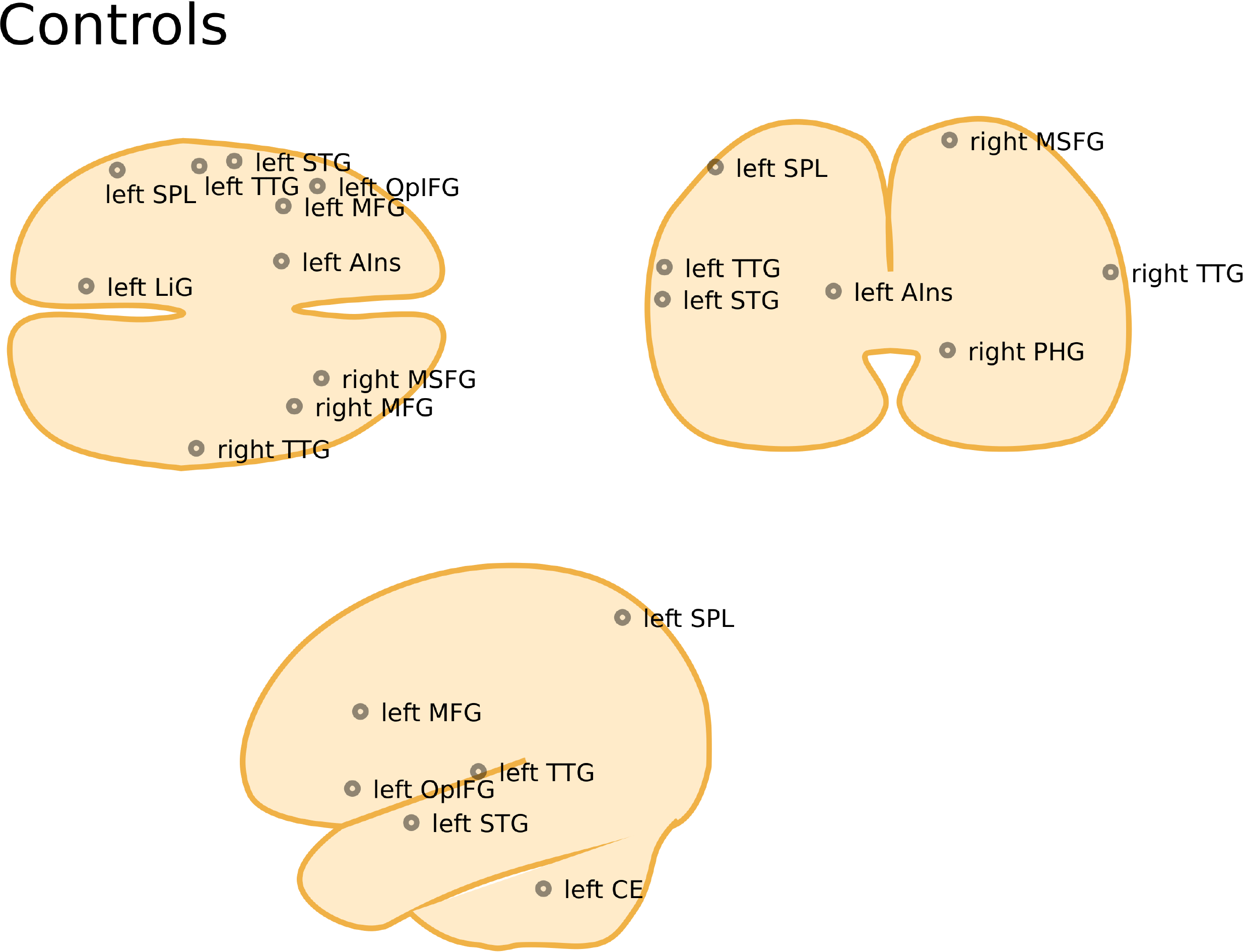}
\caption{The top nodes in representatives of loops in the distinguishing pixel birth--persistence bounds for controls.}
\label{fig:MAX_nodes_Con}
\end{centering}
\end{figure}

\begin{figure}[h!]
\begin{centering}
\includegraphics[width=\textwidth]{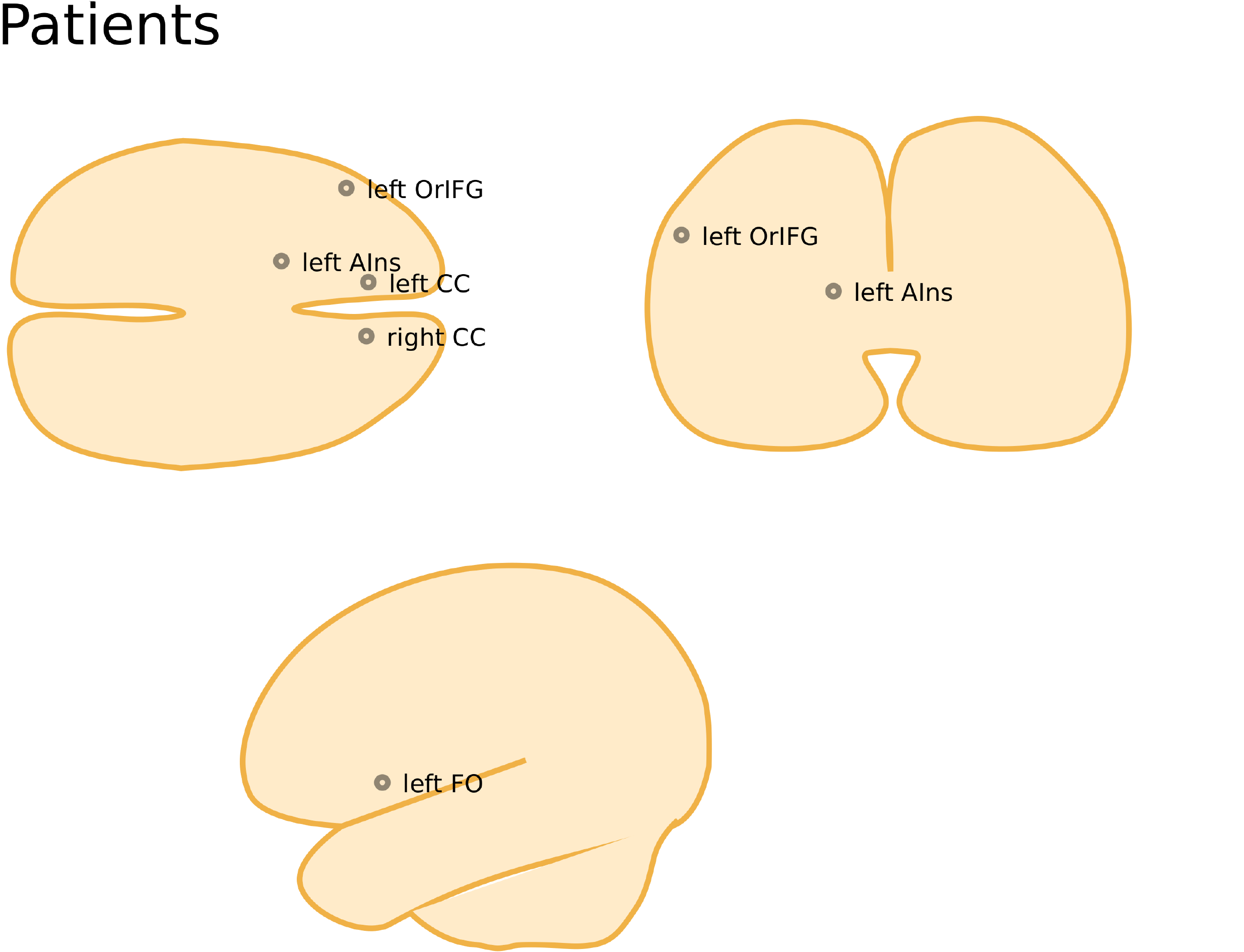}
\caption{The top nodes in representatives of loops in the distinguishing pixel birth--persistence bounds for patients.}
\label{fig:MAX_nodes_Pat}
\end{centering}
\end{figure}

In this section, we illustrate the top nodes (i.e., top brain regions) within the bounds of the distinguishing pixel birth--persistence regions for the three cohorts. Recall that each pixel in the birth--persistence plane corresponds to a bounded region of the original PD (i.e., in the \textit{birth--death} plane). We show results for siblings in Fig.~\ref{fig:MAX_nodes_Sib}, controls in Fig.~\ref{fig:MAX_nodes_Con}, and patients in Fig.~\ref{fig:MAX_nodes_Pat}.

\subsection{Supplementary Tables} \label{S1_2}

In Tables \ref{S1_Table: BrainRegionsI}--\ref{S1_Table: BrainRegionsV}, we give the numbering of the brain regions and their corresponding IDs.

\begin{table}[p]
\rotatebox{90}{\vbox{\hsize=\textheight
\caption{{\bf Node numbers (NNs) of brain regions (BRs) and their corresponding IDs (part I).}}
\begin{tabular}{ p{1cm}p{3.3cm}p{3.3cm}p{3.3cm}p{3.3cm}p{3.3cm}p{3.3cm}}
\br
NN & 1 & 2 & 3 & 4 & 5 & 6  \\ \mr
ID & 23 & 30 & 31  & 32 & 36 & 37\\ 
BR & Right accumbens area & Left accumbens area & Right amygdala & Left amygdala & Right caudate & Left caudate \\ 
\mr  
NN & 7 & 8 & 9 & 10 & 11 & 12   \\ \mr
ID &38 & 39 & 47 & 48 & 55 & 56  \\ 
BR & Right cerebellum exterior & Left cerebellum exterior & Right hippocampus & Left hippocampus & Right pallidum & Left pallidum  \\ 
\mr  
NN & 13 & 14 & 15 & 16 & 17 & 18   \\ \mr
ID & 57 & 58 & 59 & 60 & 61 & 62 \\ 
BR & Right putamen & Left putamen & Right thalamus proper & Left thalamus proper & Right ventral diencephalon & Left ventral diencephalon  \\ 
\mr 
NN & 19 & 20 & 21 & 22 & 23 & 24   \\ \mr
ID & 71 & 72 & 75 & 76 & 100 & 101 \\ 
BR & Cerebellar vermal lobules I--V & Cerebellar vermal lobules VI--VII & Left basal forebrain
 & Right basal forebrain & Right anterior cingulate gyrus & Left anterior cingulate gyrus  \\ 
\mr
NN & 25 & 26 & 27 & 28 & 29 & 30   \\ \mr
ID & 102 & 103 & 104 &105 & 106 & 107\\ 
BR & Right anterior insula & Left anterior insula & Right anterior orbital gyrus & Left anterior orbital gyrus & Right angular gyrus & Left angular gyrus  \\ 
\br  
\end{tabular}\label{S1_Table: BrainRegionsI}
}}
\end{table}

\begin{table}[p]
\rotatebox{90}{\vbox{\hsize=\textheight
\caption{{\bf Node numbers (NNs) of brain regions (BRs) and their corresponding IDs (part II).}}
\begin{tabular}{ p{1cm}p{3.3cm}p{3.3cm}p{3.3cm}p{3.3cm}p{3.3cm}p{3.3cm}}
\br
NN & 31 & 32 & 33 & 34 & 35 & 36   \\ \mr
ID & 108 & 109 & 112 & 113 &114 & 115\\ 
BR & Right calcarine cortex & Left calcarine cortex & Right central operculum & Left central operculum & Right cuneus & Left cuneus \\ 
\mr 
NN & 37 & 38 & 39 & 40 & 41 & 42   \\ \mr
ID & 116 & 117 & 118 & 119 & 120 &  121 \\
BR & Right entorhinal area &  Left entorhinal area &  Right frontal operculum &  Left frontal operculum&  Right frontal pole &  Left frontal pole \\ 
\mr  
NN & 43 & 44 & 45 & 46 & 47 &  48  \\ \mr
ID & 122 & 123 & 124 & 125 &128 & 129 \\
BR &  Right fusiform gyrus &  Left fusiform gyrus &  Right gyrus rectus &  Left gyrus rectus&  Right inferior occipital gyrus &  Left inferior occipital gyrus \\ 
\mr  
NN & 49 & 50 & 51 & 52 & 53 & 54 \\ \mr
ID & 132 & 133 & 134 &135 & 136 & 137  \\ 
BR &  Right inferior temporal gyrus &  Left inferior temporal gyrus &  Right lingual gyrus&  Left lingual gyrus &  Right lateral orbital gyrus &  Left lateral orbital gyrus \\ 
\br

\end{tabular}\label{S1_Table: BrainRegionsII}
}}
\end{table}

\begin{table}[p]
\rotatebox{90}{\vbox{\hsize=\textheight
\caption{
{\bf Node numbers (NNs) of brain regions (BRs) and their corresponding IDs (part III).}}
\begin{tabular}{ p{1cm}p{3.3cm}p{3.3cm}p{3.3cm}p{3.3cm}p{3.3cm}p{3.3cm}}
\br  
NN & 55 & 56 & 57 & 58 & 59 & 60    \\ \mr
ID & 138 & 139 &140 & 141 & 142 & 143  \\
BR &  Right middle cingulate gyrus &  Left middle cingulate gyrus &  Right medial frontal cortex &  Left medial frontal cortex&  Right middle frontal gyrus &  Left middle frontal gyrus \\ 
\mr 
NN & 61 & 62 & 63 & 64 & 65 & 66  \\ \mr
ID & 144 &145 & 146 & 147 & 148 & 149  \\ 
BR &  Right middle occipital gyrus &  Left middle occipital gyrus &  Right medial orbital gyrus &  Left medial orbital gyrus &  Right postcentral gyrus medial segment &  Left postcentral gyrus medial segment  \\ 
\mr 
NN & 67 & 68 & 69 & 70 & 71 & 72  \\ \mr
ID &150 & 151 & 152 & 153 & 154 & 155 \\
BR &  Right precentral gyrus medial segment &  Left precentral gyrus medial segment &  Right superior frontal gyrus medial segment &  Left superior frontal gyrus medial segment &  Right middle temporal gyrus &  Left middle temporal gyrus \\ 
\mr  
NN & 73 & 74 & 75 & 76 & 77 &  78   \\ \mr
ID & 156 & 157 & 160 & 161 &162 &  163 \\ 
BR &  Right occipital pole &  Left occipital pole &  Right occipital fusiform gyrus &  Left occipital fusiform gyrus &  Right opercular part of the inferior frontal gyrus &  Left opercular part of the inferior frontal gyrus  \\ 
\br

\end{tabular}\label{S1_Table: BrainRegionsIII}
}}
\end{table}

\begin{table}[p]
\rotatebox{90}{\vbox{\hsize=\textheight
\caption{
{\bf Node numbers (NNs) of brain regions (BRs) and their corresponding IDs (part IV).}}
\begin{tabular}{ p{1cm}p{3.3cm}p{3.3cm}p{3.3cm}p{3.3cm}p{3.3cm}p{3.3cm}}
\br  
NN & 79 & 80 & 81 & 82 & 83 & 84 \\ \mr
ID & 164 & 165 & 166 &167 & 168 & 169 \\
BR &  Right orbital part of the inferior frontal gyrus &  Left orbital part of the inferior frontal gyrus &  Right posterior cingulate gyrus &  Left posterior cingulate gyrus &  Right precuneus &  Left precuneus \\ 
\mr 
NN & 85 & 86 & 87 &  88 & 89 & 90 \\ \mr
ID & 170 & 171 &172 & 173 & 174 & 175 \\
BR &  Right parahippocampal gyrus &  Left parahippocampal gyrus &  Right posterior insula &  Left posterior insula &  Right parietal operculum &  Left parietal operculum \\ 
\mr
NN & 91 & 92 & 93 & 94 & 95 & 96   \\ \mr
ID & 176 &177 & 178 & 179 & 180 & 181 \\
BR &  Right postcentral gyrus &  Left postcentral gyrus &  Right posterior orbital gyrus &  Left posterior orbital gyrus &  Right planum polare &  Left planum polare \\ 
\mr  
NN & 97 & 98 & 99 & 100 & 101 & 102 \\ \mr
ID & 182 & 183 & 184 & 185 & 186 & 187\\ 
BR &   Right precentral gyrus &   Left precentral gyrus &   Right planum temporale &   Left planum temporale &   Right subcallosal area &   Left subcallosal area \\ 
\br
\end{tabular}\label{S1_Table: BrainRegionsIV}
}}
\end{table}

\begin{table}[p]
\rotatebox{90}{\vbox{\hsize=\textheight
\caption{
{\bf Node numbers (NNs) of brain regions (BRs) and their corresponding IDs (part V).}}
\begin{tabular}{ p{1cm}p{3.3cm}p{3.3cm}p{3.3cm}p{3.3cm}p{3.3cm}p{3.3cm}}
\br  
NN & 103 & 104 & 105 & 106 & 107 & 108\\ \mr
ID & 190 & 191 & 192 & 193 &194 & 195 \\ 
BR &   Right superior frontal gyrus &  Left superior frontal gyrus &  Right supplementary motor cortex &  Left supplementary motor cortex &  Right supramarginal gyrus &  Left supramarginal gyrus \\ 
\mr  
NN & 109 & 110 & 111 & 112 & 113 & 114 \\ \mr
ID & 196 & 197 & 198 & 199 & 200 & 201 \\
BR &  Right superior occipital gyrus &  Left superior occipital gyrus &  Right superior parietal lobule &  Left superior parietal lobule &  Right superior temporal gyrus &   Left superior temporal gyrus \\ 
\mr  
NN & 115 & 116 & 117 & 118 & 119 & 120 \\ \mr
ID & 202 & 203 & 204 & 205 & 206 & 207 \\
BR &  Right temporal pole &  Left temporal pole &  Right triangular part of the inferior frontal gyrus &  Left triangular part of the inferior frontal gyrus &  Right transverse temporal gyrus &  Left transverse temporal gyrus \\ 
\br  
\end{tabular}\label{S1_Table: BrainRegionsV}
}}
\end{table}

\clearpage


\end{document}